\documentclass[12pt] {article}
\usepackage[dvips]{graphicx,color}
\usepackage{rotating}

\usepackage{amssymb}

\begin{document}

\begin{titlepage}
  \enlargethispage{20pt}
\begin{minipage}[t]{.49\textwidth}
\today
\end{minipage}%
\begin{minipage}[t]{0.49\textwidth}

\end{minipage}
  \begin{center}
    \noindent{\large \bf

 PROPOSAL FOR AN EXPERIMENTAL PROGRAM\\ IN NEUTRINO PHYSICS
AND PROTON DECAY\\ IN THE HOMESTAKE LABORATORY

    }
  \end{center}

\begin{center}

%DRAFT DRAFT DRAFT 

\end{center}

%\begin{figure}[htbp]
%  \begin{center}
%  \includegraphics*[width=\textwidth]{us-map.eps}
%  \end{center}
%\end{figure}

\newpage

\pagestyle{empty}

This document contains figures in color.

\newpage

\noindent{\large \bf List of Collaborators}

\begin{center}

 M. Diwan, S. Kettell,  
L. Littenberg, W. Marciano, Z. Parsa, N. Samios, S. White \\
{\sl Department of Physics, 
  Brookhaven National Laboratory, Upton, NY }
\smallskip

R.~Lanou \\
{\sl Department of Physics,  Brown University, Providence, RI}

W.~Leland, K.~Lesko \\
{\sl Department of Physics, 
University of California, Berkeley, CA}

Karsten Heeger, W. Y. Lee \\
{\sl Lawrence Berkeley Laboratory, Physics Division, 
Berkeley, CA}

W.~Frati, K.~Lande, A.~K.~Mann,
R.~Van Berg  \\
{\sl
Department of Physics and Astronomy, 
  University of Pennsylvania
  Philadelphia, PA }

K.~T.~McDonald \\
{\sl Department of Physics, Princeton Univesity, Princeton, NJ}

D.~B.~Cline \\
{\sl Department of Physics and Astronomy, 
University of California, Los Angeles, CA  }

P.~Huber, V.~Barger \\
{\sl Department of Physics, University of Wisconsin, Madison, WI }

D.~Marfatia \\
{\sl Department of Physics and Astronomy, 
 University of Kansas, Lawrence, KS}

T. Kirk \\
{\sl Department of Physics, University of Colorado,
Boulder, CO}

Renato Potenza \\
{\sl Instituto Nazionale di Fisica Nucleare,
  Dipartimento de Fisica e Astronomia,
  Universita di Catania,
  64, Via S. Sofia,
  I-95123 Catania,
  Italy}

\end{center}
\end{titlepage}

\abstract{  
This report is intended to describe first, the principal physics reasons for an
ambitious experimental program in neutrino physics and proton decay based on 
 construction of a series of massive water Cherenkov detectors located
deep underground (4850 ft) in the Homestake Mine 
of the South Dakota Science and Technology Authority (SDSTA); and  second,
the engineering design of the underground chambers to house the Cherenkov
detector modules; and third, the conceptual design of the water Cherenkov
 detectors themselves for this purpose.
Included in this document  are preliminary 
costs and time-to-completion estimates
which have been exposed to  acknowledged experts in their respective areas.
We have included  some
contingency factors. Nevertheless, we recognize that 
much more extensive documentation and contingency estimates will be needed
for a full technical design report.

In this proposal we show the event rates and 
physics sensitivity for 
beams from both FNAL 
(1300 km distant from Homestake) and BNL (2540 km distant from Homestake).
 The program we propose will benefit with
 a beam from FNAL because of the high intensities
currently available from the Main Injector with 
modest upgrades.  The possibility of tuning the primary 
 proton energy
over a large range from 30 to 120 GeV  also 
 adds considerable flexibility to 
the program from FNAL.  

}

\newpage

\newpage

\pagestyle{plain}

\pagenumbering{roman}

\tableofcontents

% \vspace{.2in}

\newpage

\pagenumbering{arabic}

\vspace{2ex}

\section*{Preface}

\vspace{2ex}

This report is intended to describe first, the principal physics reasons for an
ambitious experimental program in neutrino physics and proton decay based on 
 construction of a series of massive water Cherenkov detectors located
deep underground (4850 ft) in the Homestake Mine 
of the South Dakota Science and Technology Authority (SDSTA); and  second,
the engineering design of the underground chambers to house the Cherenkov
detector modules; and third, the conceptual design of the water Cherenkov
 detectors themselves for this purpose.

In this proposal we show the event rates and 
physics sensitivity for 
beams from both FNAL 
(1300 km distant from Homestake) and BNL (2540 km distant from Homestake).
 The program we propose will benefit with
 a beam from FNAL because of the high intensities
currently available from the Main Injector with 
modest upgrades.  The possibility of tuning the primary 
 proton energy
over a large range from 30 to 120 GeV  also 
 adds considerable flexibility to 
the program from FNAL.  
On the other hand the  beam from BNL over the larger distance will produce 
very large matter effects, and consequently    a hint of new physics 
(beyond CP violation)  can be better tested with that configuration. 
  In this proposal we focus on the 
CP violation physics.

Included in this document  are preliminary 
costs and time-to-completion estimates
which have been exposed to  acknowledged experts in their respective areas.
This presentation is not, however, to be taken as a technical design report
with the extensive documentation and contingency costs that a TDR usually
entails. Nevertheless, some
contingency factors have been included in the estimates given here.

The essential ideas expressed here were first laid out in a letter of
 intent to the interim director of the Homestake Laboratory on July 26, 2001. 
 Since that time, the prospect of a laboratory in the Homestake Mine
 has been realized, and the design of a long baseline neutrino experiment
 has been refined. 
 The extrapolation contained in this proposal is within the common domain 
of thinking in the area of physics discussed here.  It needs now only the 
encouragement of the funding agencies, NSF and DOE.

\vspace{2ex}

\section{Introduction}

\label{intro}

\vspace{2ex}

With the discovery of neutrino oscillations, it becomes 
 realistic to test for a violation of CP-invariance in the lepton sector
of elementary particles. As is well-known, CP-invariance in the hadron
sector of elementary particles was discovered more than 40 years ago and has
 been the subject of study---experimental and theoretical---since that time. 
In the event that such  violation is found and measured in the lepton sector, 
the implications of that finding would be profound for elementary particle
physics and similarly for astrophysics and cosmology. This is the principal
goal of the long-baseline neutrino program in the Homestake Laboratory.

We have previously pointed out that 
for a given resolution on $\delta_{CP}$ the total 
number of neutrino events needed,
 which determines
the detector size and beam intensity, is approximately independent of
the baseline length as well as the value of $\sin^2 2 \theta_{13}$ 
\cite{ref3, vlbl}, 
as long as the electron signal is not
background dominated or the value of $\theta_{13}$ too small 
($\sin^2 2\theta_{13}<0.003$\cite{ref4}).   
Therefore this very  long baseline ($\stackrel{ >}{\sim}$ 1300 km) program
is attractive regardless of other projects in neutrino physics 
because it provides   high sensitivity to 
$\theta_{13}$, CP violation, and the mass hierarchy as well as 
 access to much richer physics phenomena. 

There are three additional goals that can be addressed by the same
detector at the same time: (1) high precision studies of neutrino
oscillations,  (2) a search for nucleon decay 
 to yield a
substantially improved sensitivity to the decay rate $p \rightarrow
e^+ + \pi^0$, and perhaps to other decay modes, and (3) a search
for supernova neutrino bursts. All of these goals
have been extensively discussed in the literature and are recognized
as important issues to be determined experimentally in the progress of
elementary particle physics.

\begin{figure}[htbp]
  \begin{center}
  \includegraphics*[width=\textwidth]{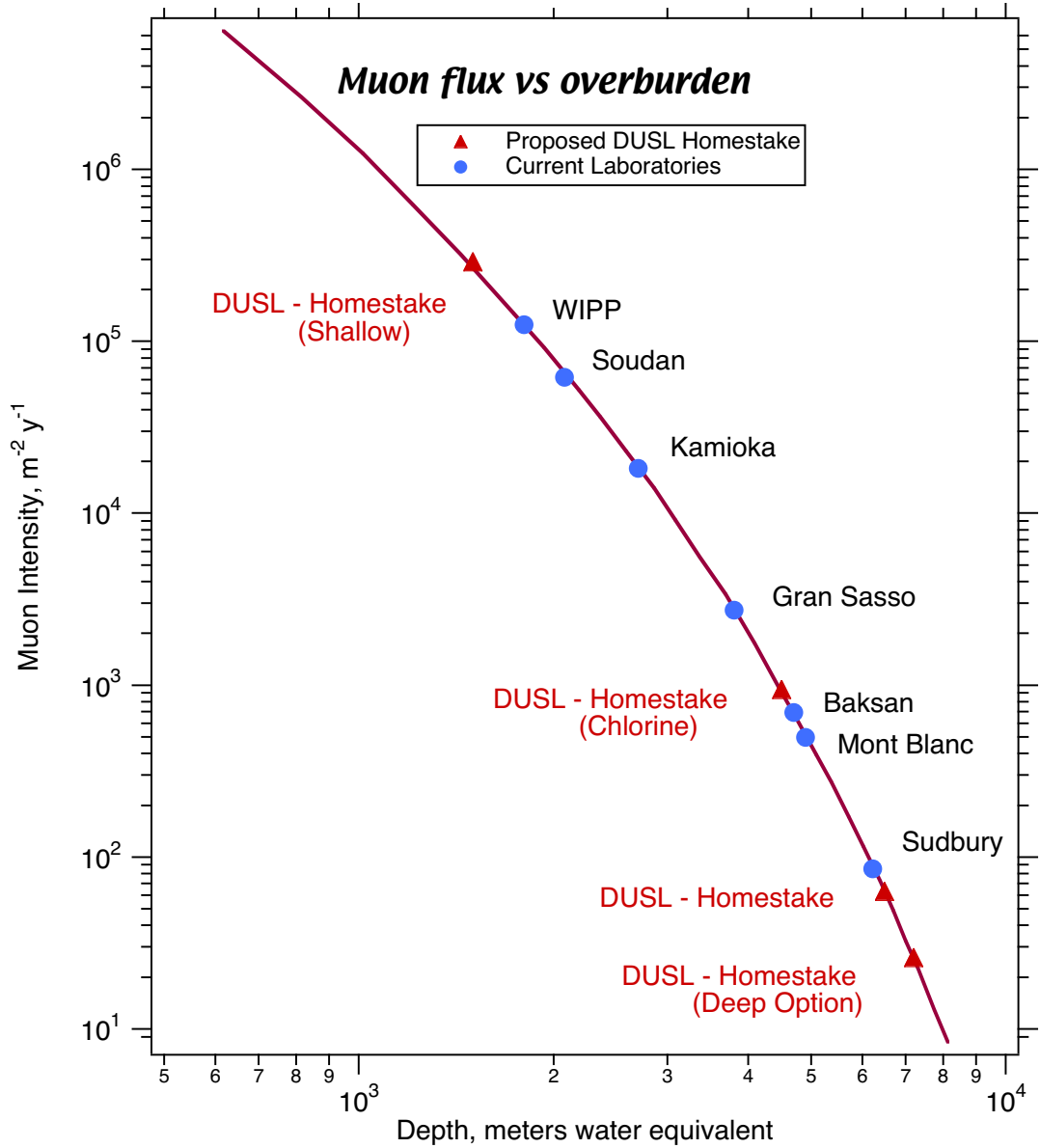}
  \caption[Muon rate versus depth]{
Cosmic ray muon intensity as a function of depth in meters water 
equivalent at the Homestake based 
Deep Underground Science and Engineering Laboratory (DUSEL). The depth and rates at other 
underground laboratory locations in the world are also indicated for 
comparison. Various locations in Homestake of potential sites for 
physics experiments are indicated by a red triangle.  
The Homestake (Chlorine) depth (4850 ft.)  is the potential 
location for the detector proposed in this document.  The rate of muons for 
a 100 kT detector at this depth is approximately 0.1 Hz.  

}
   \label{fig1}
  \end{center}
\end{figure}

The design of the detector for this program may be briefly stated.
The three goals described here require high statistical sensitivity
and good discrimination against backgrounds; these in turn determine
the size, the type and nature of the detector. 
Considering the present intensity levels available for neutrino beams from US 
accelerator laboratories (Fermi National Accelerator Laboratory (FNAL)  
or Brookhaven National Laboratory (BNL)),   
a detector fiducial
mass of $\geq$ 300 ktons is essential to acquire the statistical precision
of the test for CP invariance and to study neutrino oscillations at the 
statistical level of 1--3\%.
 Similar detector mass is
 also essential to extend the limit
on the nucleon decay lifetime. To achieve such large mass and simultaneously
obtain sufficient 
 information about  observed events for effective background elimination
 leads to  the choice of detector as a water
Cherenkov detector. To make it possible to use the three detector modules in
concert or independently specifies their modularity, 
and finally to protect  them
from the cosmic ray flux demands that they be operated deep underground.
This last condition is amply satisfied at Homestake as shown in
 Fig. \ref{fig1}.

It is also worthwhile in this brief introduction to complete the
description of the long baseline neutrino program in the Homestake
Laboratory by addressing two misconceptions.

First is the issue of what is referred to as ``extreme cost'' of such
a large mass neutrino detector and the impact of that cost on other
research plans for expensive facilities.  The modular detector
proposed here will require four or five years to construct from the
time funds become available. Although the chambers for all
three modules may be excavated at the same time, for time and cost
savings, it is likely that the detector modules will be finished
serially. As a result, the first module can be shaken down and begin
operating while construction of the others is being completed. This
means an early start of data-taking and control of the rate at which
funds are expended if such fine-tuning were to be requested by the
funding agency. In any event, the present estimate of the cost of the
complete 3 module detector, which will be presented in detail later in
this report, is \$308M, or \$62M per year over  the five years of
construction.

This should  be regarded as a well considered 
 allocation of resources
 since the useful life 
of the detector complex will extend over a period of at least ten years
following the onset of data-taking, and very likely much 
 longer without the beam  to continue study of atmospheric neutrinos,
  neutrino emissions
from Type 2 supernovae, and proton decay.

Second, the CP-test needs a broad band neutrino
beam from a high energy accelerator 
to be discussed briefly below, but is properly the province of
FNAL and BNL which are now involved in discussions relating to the
technical aspects of that beam and its delivery to Homestake
Laboratory (distances: $\sim$ 1290 km 
from FNAL and $\sim$ 2540 km from BNL). 
A design for such beam from BNL exists and has been well documented.   
We hope that as part of the FNAL/BNL study on beams and accelerator 
intensity a design for such a beam will be produced at FNAL.
It is clear even from initial studies that due to the high intensities 
available at the Main Injector and the flexibility in proton 
energy over a wide range (30 to 120 GeV), a beam from FNAL could be 
very advantageous.  
We hope that a new design will produce adequate  
intensity 
neutrino beam  for the initial CP-test without a need for large 
upgrades to the proton complex,  but such upgrades will be 
desired subsequently.

 It is expected that
construction of a beam at either accelerator would take about four or
five years and coincide approximately with the schedule for the
Homestake detectors if work were to begin at both projects at the same
time. It should be emphasized, however, that more than a rough
coincidence in time is not demanded; as discussed below, there is
important physics to be studied in the Homestake detector without a
neutrino beam, provided the beam becomes available within two
years after completion of the full detector.

We turn now to the second section of this report: an abbreviated
description of neutrino physics and CP-violation, and proton decay.

\section{Neutrino Oscillations and CP-Invariance Violation}

\begin{figure}[htbp]
  \begin{center}
 \includegraphics*[width=0.5\textwidth]{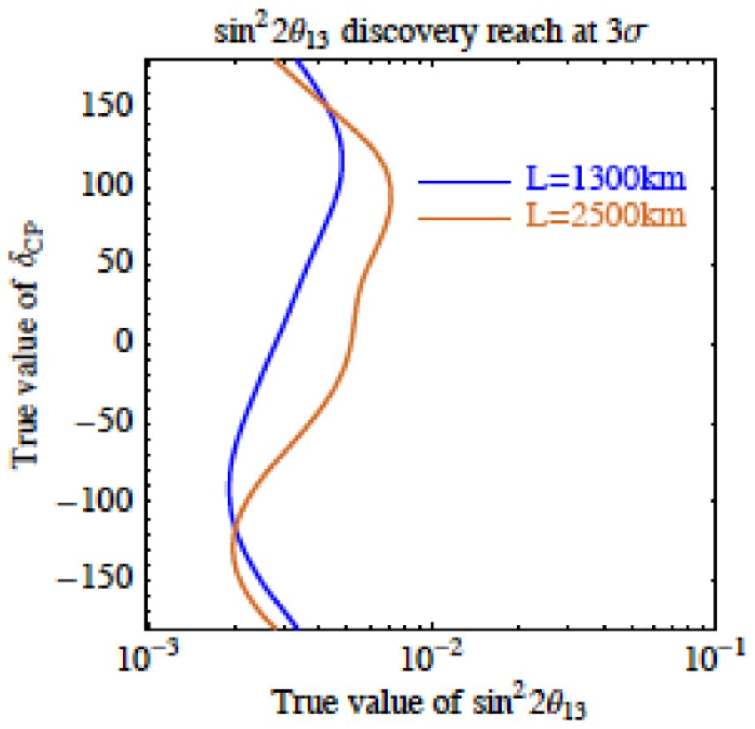}
 \includegraphics*[width=0.5\textwidth]{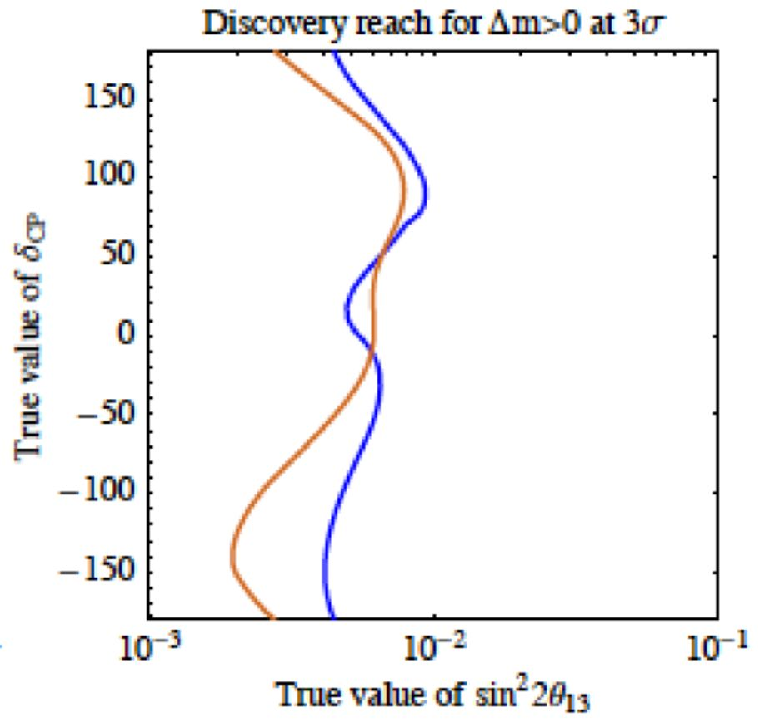}
  \caption[Physics reach of broadband beam to Homestake]{
This graph shows  the 3 sigma physics reach for 
$\sin^2 2 \theta_{13}$ (top),
 and resolution of the mass hierarchy (bottom) of the 
program we are proposing here. The region to the right of the curves will be 
excluded for $\sin^2 2 \theta_{13}$ (top) in case of no signal. 
If there is a signal then 
 the mass hierarchy will be determined to the 
right hand side of the curves in the bottom plot. 
 The calculation was made for the
exposure of $300 kT\times 1 MW\times 
5 yrs$ each for neutrino and 
anti neutrino running. The assumptions about the spectrum 
and proton 
energy are 
explained in the text. We assume running time of $1.7\times 10^7$ sec per year.
The reach for both 1300 km and 2540 km baselines is shown. 
}
   \label{fig2}
  \end{center}
\end{figure}

\begin{figure}[htbp]
  \begin{center}
 \includegraphics*[width=0.5\textwidth]{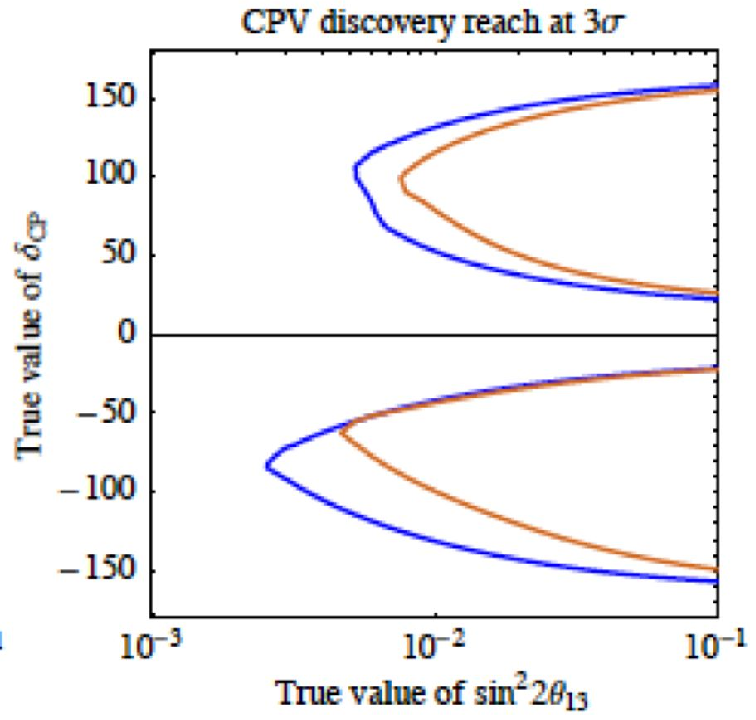}
  \caption[CP violation reach of broadband beam to Homestake]{
This graph shows  the 3 sigma  reach for detection of CP violation 
as a function of $\sin^2 2 \theta_{13}$ and $\delta_{CP}$,
There will be a 3 sigma signal for CP violation in 
 the region to the right of the curves. 
The CP parameter $\delta_{CP}$ will be determined to at least 
3 sigma with no ambiguities in a simultaneous determination (or
fit) that also yields $\theta_{13}$ and the mass hierarchy. 
The assumptions for the detector exposure are the same as 
in Fig. \ref{fig2}. 
}
   \label{figcp}
  \end{center}
\end{figure}

The physics reach of the program here in the study of neutrino-oscillations
and CP-violation is extensive (see Fig. \ref{fig2} and Fig. \ref{figcp})\cite{huber, globes}
 and needs to be presented in quantitative
detail to be appreciated. A compact but complete presentation of that
material, ``Very Long Baseline Neutrino Oscillation Experiments
for Precise Measurements of Mixing Parameters and CP Violating Effects'',
 is published in Phys. Rev. D68:012002 \cite{vlbl}  and also in 
eprint hep-ph/0407047 \cite{hql04}. We summarize here  parts of
the introduction and conclusions of those reports. We will also use the 
materials generated in recent workshops on the subject as well as  
 the reports from the NSF DUSEL 
(Deep Underground Science and Engineering Laboratory) 
initiatives. 

The first violation of CP-invariance was observed in
 the $K^0 - \bar{K}^0$ system about forty years ago. Until that experiment, 
it was thought that
the product of those  symmetry factors was invariant and preserved
the symmetry between a fundamental neutral particle and its antiparticle.
Recently,  CP violation has been observed in a second quark system, the
$B^0 - \bar{B}^0$ system. Following that, interest is raised in the question
of the presence of CP-invariance violation in the neutrino sector.

The observation of neutrino oscillations opens the neutrino sector as a
promising region in which a CP asymmetry between neutrino and
antineutrino might be exhibited. 
This would fuel further speculation that a primordial universal
asymmetry in the early universe accounts for the matter-antimatter asymmetry
we observe now. The suggestion that a long baseline neutrino oscillation
experimental program would be a fertile region to explore for that purpose
has been studied intensively 
(e.g., see 
\cite{NUSEL, freund, barger, marfatia1, geer} and references therein)
 in the last few years as a result of the
experimental data on neutrino oscillations, but a conclusive demonstration of
neutrino oscillations and a search for leptonic 
CP-invariance violation awaits a more
ambitious experimental program such as the one proposed here.

Measurements of solar and atmospheric neutrinos have provided strong evidence
for non-zero neutrino masses and mixing \cite{ref1, sk, kamland, sno, k2k}. 
Interpretation of the
experimental results is based on oscillations of one neutrino flavor state,
$\nu_e$, $\nu_{\mu}$, or $\nu_{\tau}$, into the others, and described quantum
mechanically in terms of neutrino mass eigenstates, $\nu_1$, $\nu_2$, and
$\nu_3$. The mass squared differences involved in the transitions are measured
to be approximately $\Delta m^2_{21} \equiv {m (\nu_2)}^2 - {m(\nu_1)}^2 
\approx 8 \times 10^{-5}eV^2$ (solar parameters)  and
$\Delta m^2_{32} \equiv {m(\nu_3)}^2 - {m (\nu_2)}^2 \approx 
\pm 2.5 \times 10^{-3}eV^2$ (atmospheric parameters),
 with large mixing strengths,
 $\sin^2 2 \theta_{12} \approx 0.86$ and $\sin^2 2 \theta_{23} \approx 1.0$ 
in both cases (See Table \ref{pars}). 
The  currently favored parameters
open the possibility for an accelerator based  long baseline 
program that can measure both solar and atmospheric oscillation
parameters in a single experiment with high precision, and search 
for CP violation and new physics.

\begin{figure}[htbp]
  \begin{center}
  \includegraphics*[width=0.5\textwidth]{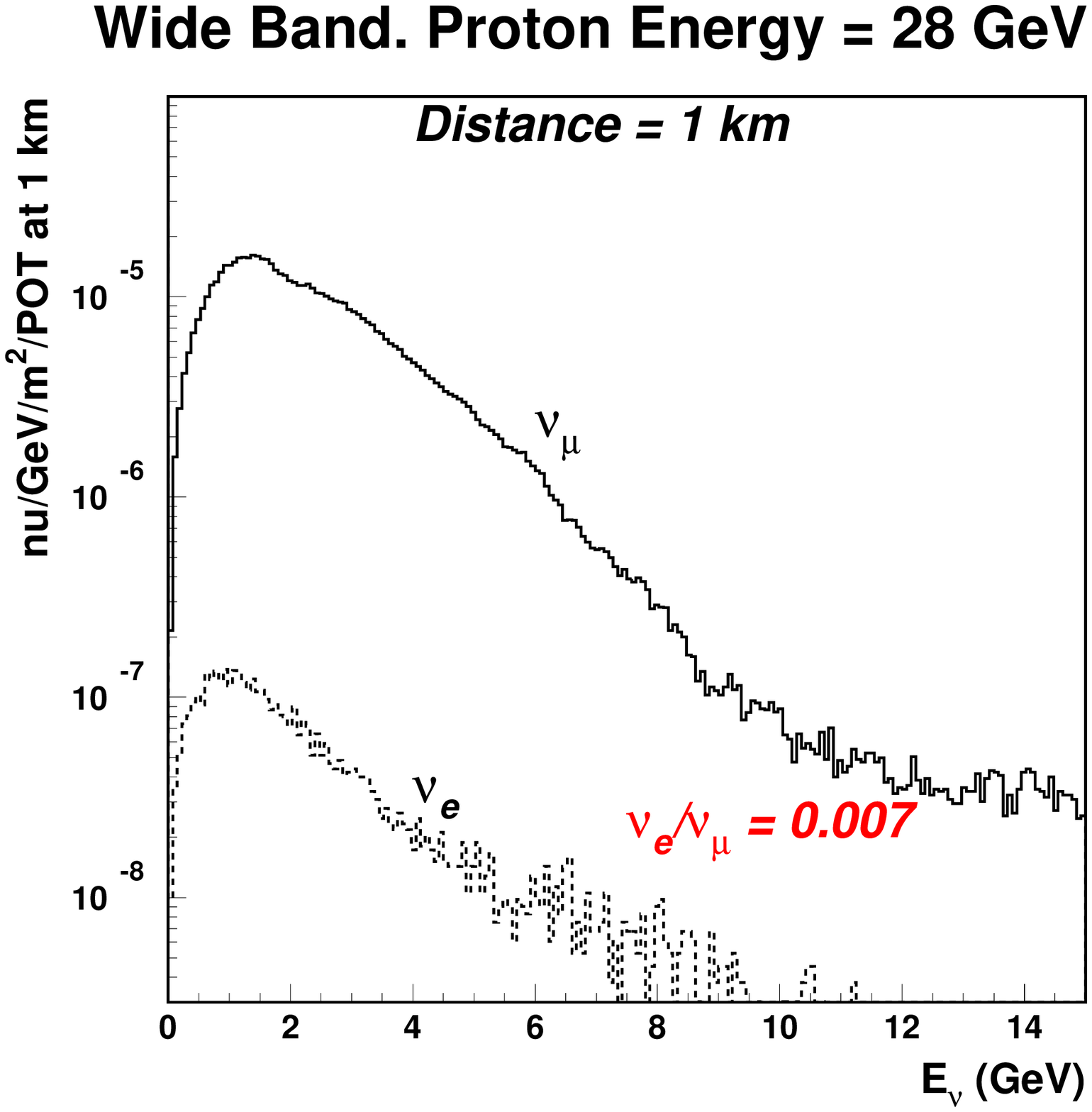}
  \includegraphics*[width=0.5\textwidth]{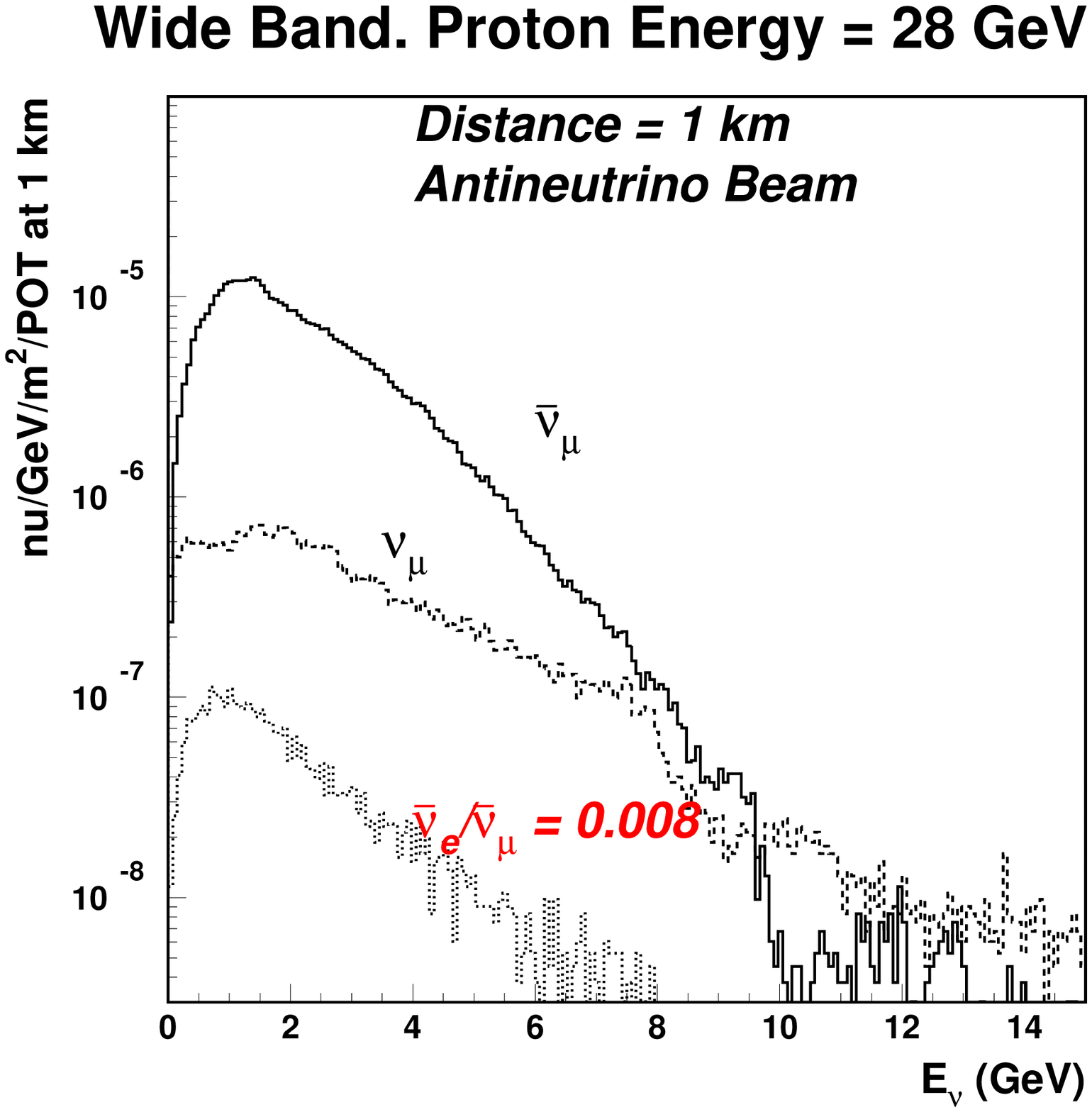}
  \caption[Neutrino spectra]{
This plot shows the spectrum of neutrinos (top) and anti-neutrinos 
(bottom)  from 
28 GeV protons with a wide band horn design (from BNL) calculated using 
a GEANT based Monte Carlo program. The decay tunnel 
diameter was assumed to be 4 meters and length to be 180 Meters.
Such a beam could originate from FNAL or BNL. At FNAL, the proton beam energy 
could be higher and the tunnel length could be longer, both yielding 
a more intense beam per proton; this optimization is in progress.   
}
   \label{spectra}
  \end{center}
\end{figure}

We have simulated and analyzed a feasible long baseline neutrino
oscillation program consisting of a low energy, wide band neutrino beam
produced by ${\sim}$ 1 MW of 28 to 60
GeV protons incident on a carbon target with magnetic
horn focusing of pions and kaons and a large water Cherenkov 
 detector at a distance
of 1300 or 2500 km from the neutrino source. 
The results reported in \cite{vlbl} were based on a fast 
simulation of the signal and background events 
 based on a Monte Carlo that included a parameterized response of 
a water Cherenkov detector. Since then a study has been performed with complete
Monte Carlo simulation based on Super Kamiokande
 (reported by Chiaki Yanagisawa \cite{chiaki})  using the spectra shown in 
Fig. \ref{spectra}. We have adjusted our detector parameterization for the 
fast simulation to conform to this new study. The plots shown in this 
proposal are largely based on this. 
The  neutrino beam with a total intensity of about 
$4.7 \times 10^{-5} \nu/m^2/POT$ using 28 GeV protons 
at a distance of 1 km from the target 
could be provided by either the FNAL Main Injector
\cite{foster, fnalmarch, mcginnis} or the BNL-AGS \cite{agsup}. 
If sited at FNAL, optimization of the beam energy from 30 to 120 GeV and
choice of decay tunnel length up to 400 meters is possible. This work is 
on-going and could result in better performance of the beam intensity.

This single  program  can produce measurements of
all parameters in the neutrino mixing matrix through observation 
of the disappearance channel,
$\nu_{\mu} \rightarrow \nu_{\mu}$, and the appearance channel
 $\nu_{\mu} \rightarrow \nu_e$. The project  is also sensitive 
to the mass ordering of
neutrinos using the observation of the matter effect in the appearance
channel  through the currently unknown parameter $\sin^2 2 \theta_{13}$.
Nevertheless, the program is intended primarily to measure the strength
of CP invariance violation in the neutrino sector and will provide a
measurement of the CP phase, $\delta_{CP}$, or alternatively the CP violating
quantity, $J_{CP}$, if the one currently unknown neutrino oscillation
mixing parameter $\sin^2 2 \theta_{13} \geq 0.01$, a value about 15 times
lower than the present experimental upper limit. 
We point out that 
for a given resolution on $\delta_{CP}$ the number of neutrino events needed,
 which determines
the detector size and beam intensity, is approximately independent of
the baseline length as well as the value of $\sin^2 2 \theta_{13}$ 
\cite{ref3}, 
as long as the electron signal is not
background dominated or the value of $\theta_{13}$ too small 
($\sin^2 2\theta_{13}<0.003$\cite{ref4}).   
Therefore this very  long baseline ($\stackrel{ >}{\sim}$ 1300 km) program
is attractive regardless of other projects in neutrino physics 
because it provides   high sensitivity to 
$\theta_{13}$, CP violation, and the mass hierarchy as well as 
 access to much richer physics phenomena. 
 Lastly, we comment that most of  the rich
physics program including 
a preliminary search for 3-generation model 
CP effects through spectral 
distortions can be carried out
 by neutrino running alone. Nevertheless, 
anti-neutrino data will be essential to establish CP violation 
and to distinguish new physics in the leptonic sector from the 3-generation 
model. 
The shape of the disappearance and appearance signals over multiple 
oscillations from neutrino and anti-neutrino running  can be used to constrain 
the magnitude of effects of new physics
from, e.g., sterile neutrino mixing, extra dimensions, exotic interactions
with matter, etc.\cite{marfatia}.

\begin{table} 
\caption[Currently known neutrino oscillation parameters] 
{In this table we list the best known neutrino oscillation
parameters from a recent global fit\cite{strumia}. Other  fits give 
similar results.} 
\begin{center} 
\begin{tabular}{|l|r|} 
\hline 
Parameter & Value  \\
\hline  
$\Delta m^2_{21}$ & $(8.0\pm0.3)\times 10^{-5} eV^2$ \\
$\sin^2 2\theta_{12}$ & $0.86\pm 0.04$ \\
\hline 
$|\Delta m^2_{32}|$ & $(2.5\pm0.3)\times 10^{-3} eV^2$ \\
$\sin^2 2 \theta_{23}$ & $1.02\pm0.04$ \\
\hline  
$\sin^2 2 \theta_{13}$ & $<0.12$ \\
$\delta_{CP}$ & Unknown \\
\hline
\end{tabular} 
\end{center} 
\label{pars} 
\end{table}

The FNAL main injector has the ability to produce 
proton beams from $\sim 30 $ GeV to $120$ GeV. 
It is been shown that when the Tevatron completes 
operations around 2010, a number of improvements can be 
made to the FNAL complex to raise the total power 
of the main injector to 1.1 MW at 120 GeV \cite{fnalmarch}. 
There are studies to find 
 more improvements to maintain the power level at lower 
energies as well \cite{foster, mcginnis}. 
For the BNL-AGS a plan exists to improve the facility to 1 MW 
beam power at 28 GeV \cite{agsup}. 
For the purposes of this proposal we assume that a beam can be
made with 28 GeV protons with a power level of 1 MW.
The spectrum for neutrinos and antineutrinos 
is shown in Fig. \ref{spectra}.   
Optimization of the beam spectrum  is in progress as part of the joint effort 
between BNL and FNAL. In particular, 
it is clear that higher energy protons will 
yield a more intense and broader neutrino flux per proton. This has to be 
balanced against the need to suppress neutral current backgrounds from 
higher energy neutrinos.  

The  neutrino event rate for the exposure assumed for Fig. \ref{fig2} 
is approximately 53000 (14000) events per year for the FNAL (BNL) 
to Homestake distance of 1300 km (2540 km). The anti-neutrino event 
rate is  16000 (4200) events per year for the FNAL (BNL) distance. 
Total exposure was assumed to be 5 years each.  The anti-neutrino
event yields are smaller due to the smaller cross section. Therefore 
we anticipate that a further factor of 2 upgrade of the beam intensity 
will be in the plans after the initial 5 year neutrino exposure. 
Nevertheless,
we have not assumed such an upgrade in the calculations of Fig. \ref{fig2}. 
 
\begin{figure}[htbp]
  \begin{center}
 \includegraphics*[width=0.5\textwidth]{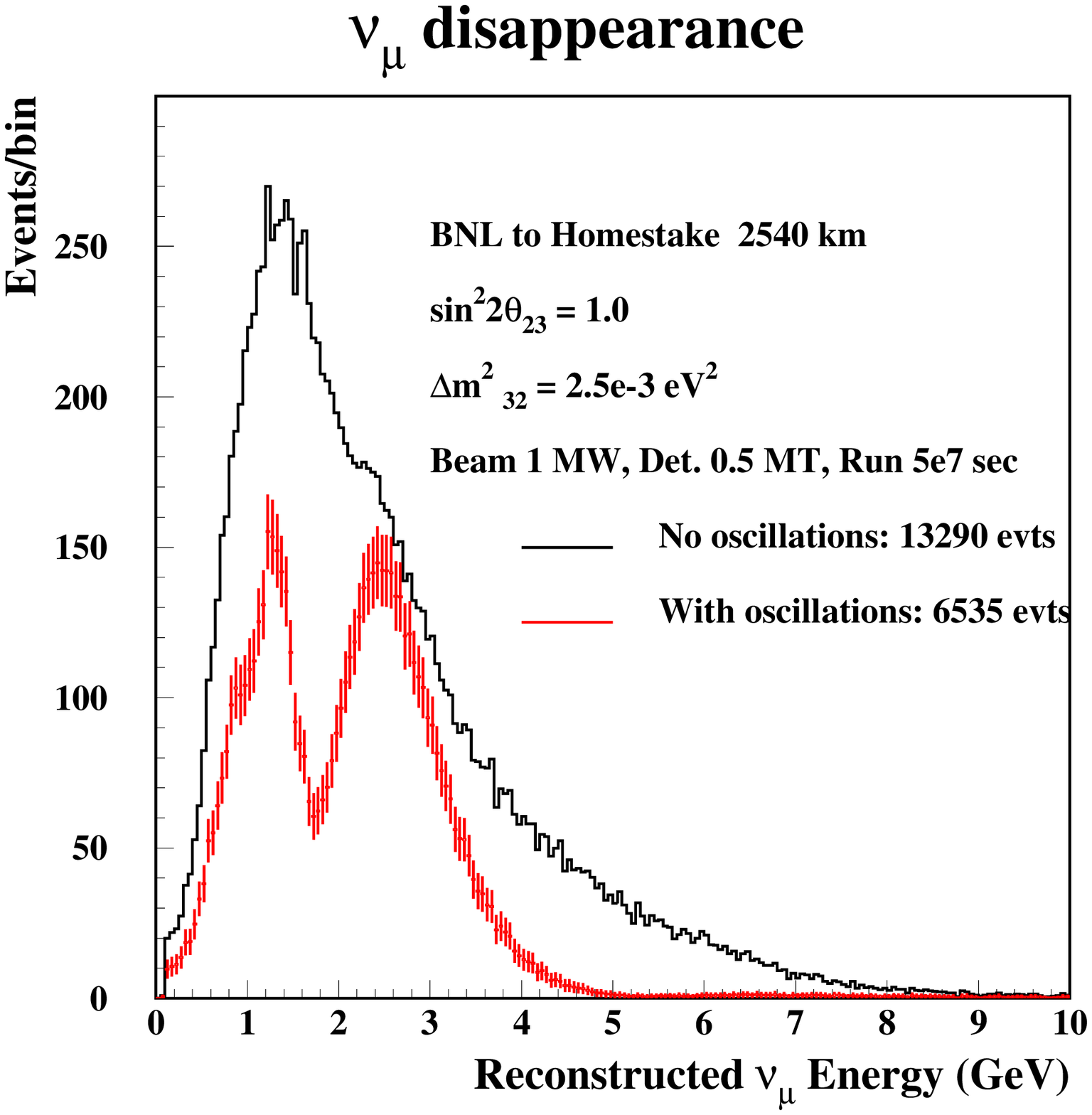}
  \includegraphics*[width=0.5\textwidth]{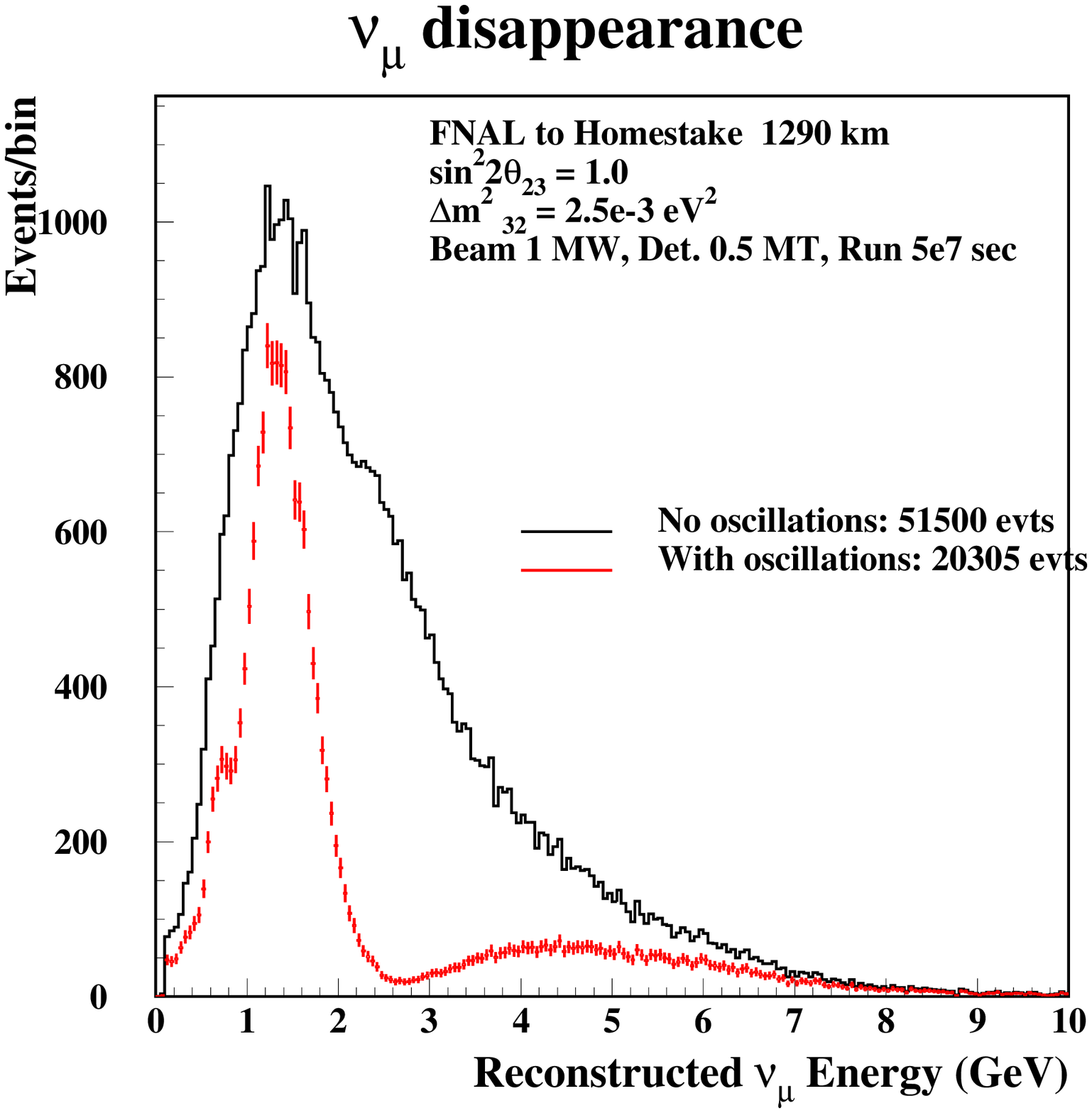}
  \caption[Spectra for charged current $\nu_\mu$ interactions]{
Charged current $\nu_\mu$ interaction spectrum after cuts to select 
clean single muon events. The plot includes background from events 
in which there are additional charged pions below Cherenkov threshold. 
The plots with and without oscillations  are shown for 2540 (top) and 
1300 (bottom)  km
distances.
}
   \label{disap}
  \end{center}
\end{figure}

\begin{figure}[htbp]
  \begin{center}
 \includegraphics*[width=0.5\textwidth]{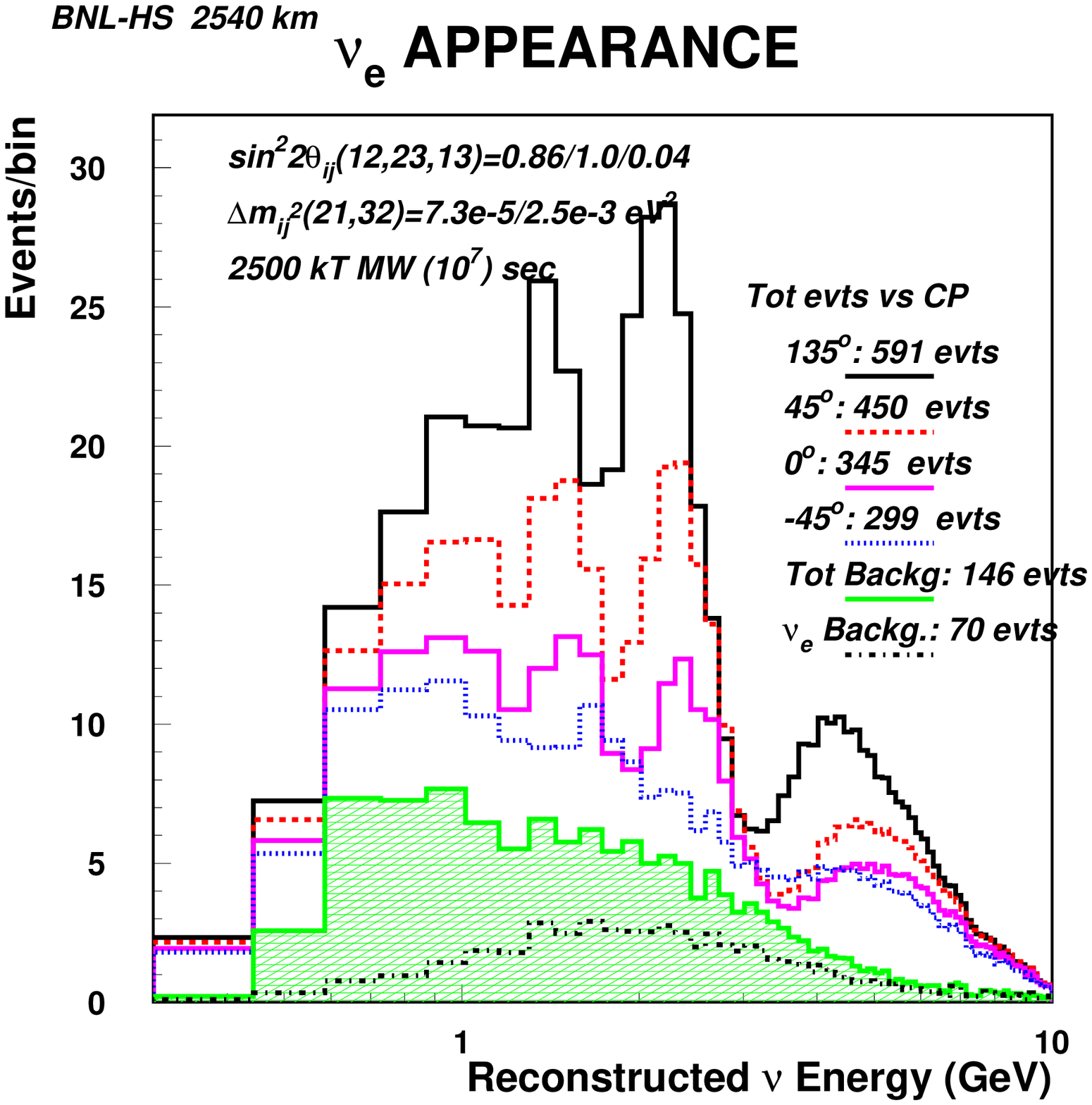}
  \includegraphics*[width=0.5\textwidth]{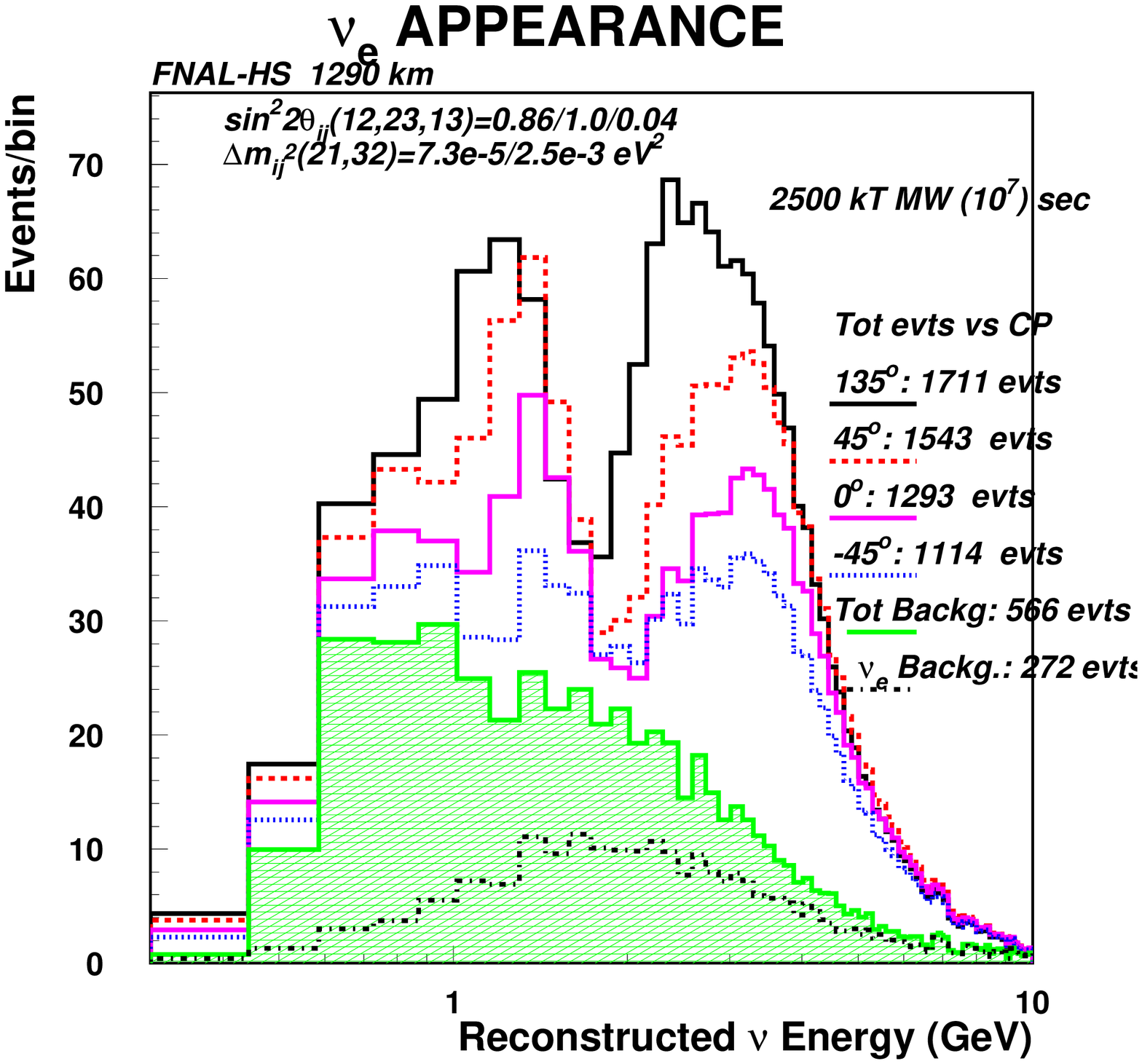}
  \caption[Spectra for $\nu_e$ appearance]{
Charged current $\nu_e$ interaction spectrum after cuts to reduce background.
The oscillation parameters assumed are indicated in the figures. We have used the value of $\sin^2 2 \theta_{13} =0.04$ for illustrative purposes.  
The CP phase is changed for the various curves as indicated in the figures. 
For these figure  the mass hierarchy is assumed to be "normal". 
The spectra  are shown for 2540 (top)and 1290 (bottom) km
distances.
}
   \label{app}
  \end{center}
\end{figure}

\begin{figure}[htbp]
  \begin{center}
 \includegraphics*[width=0.5\textwidth]{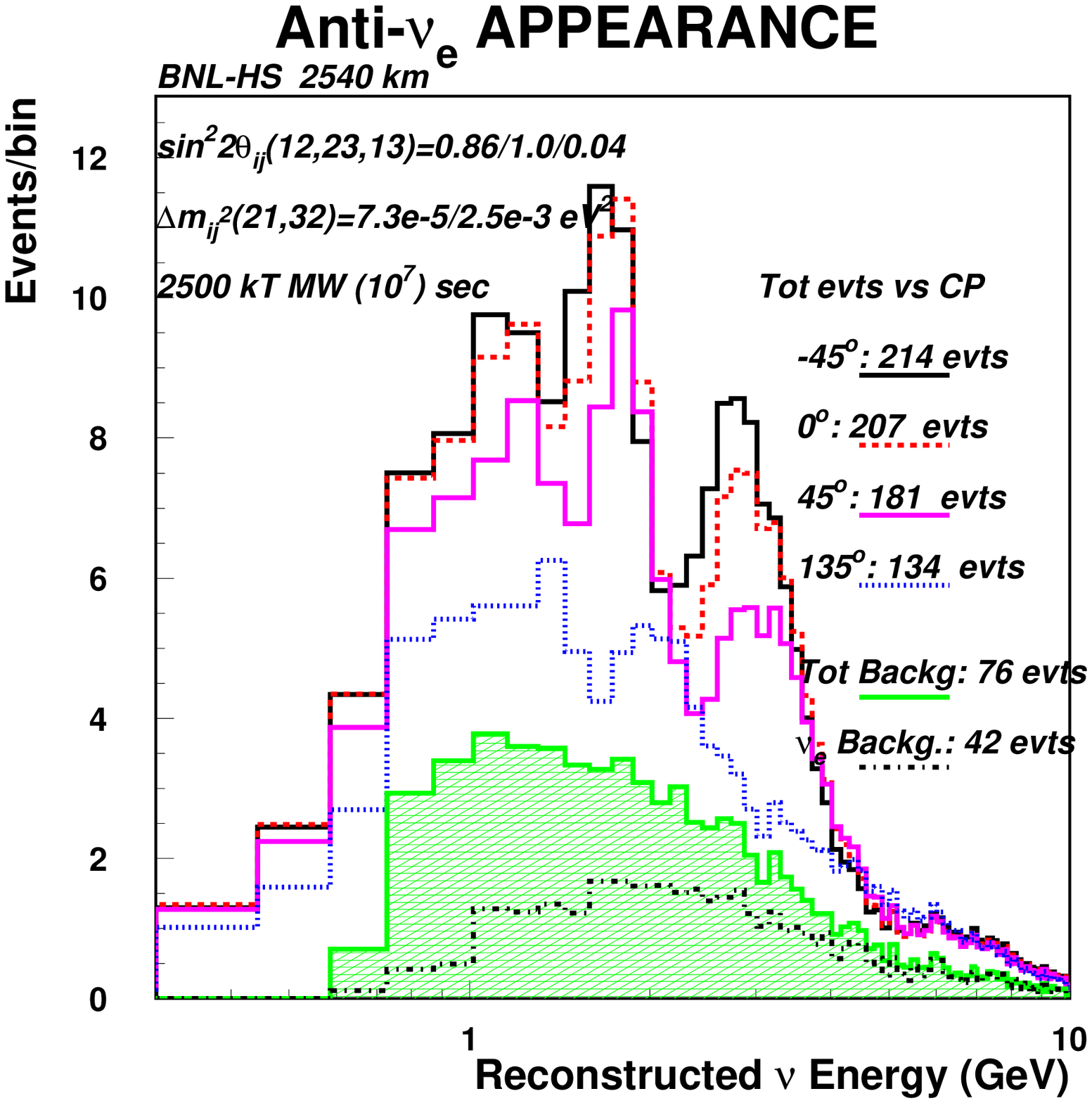}
  \includegraphics*[width=0.5\textwidth]{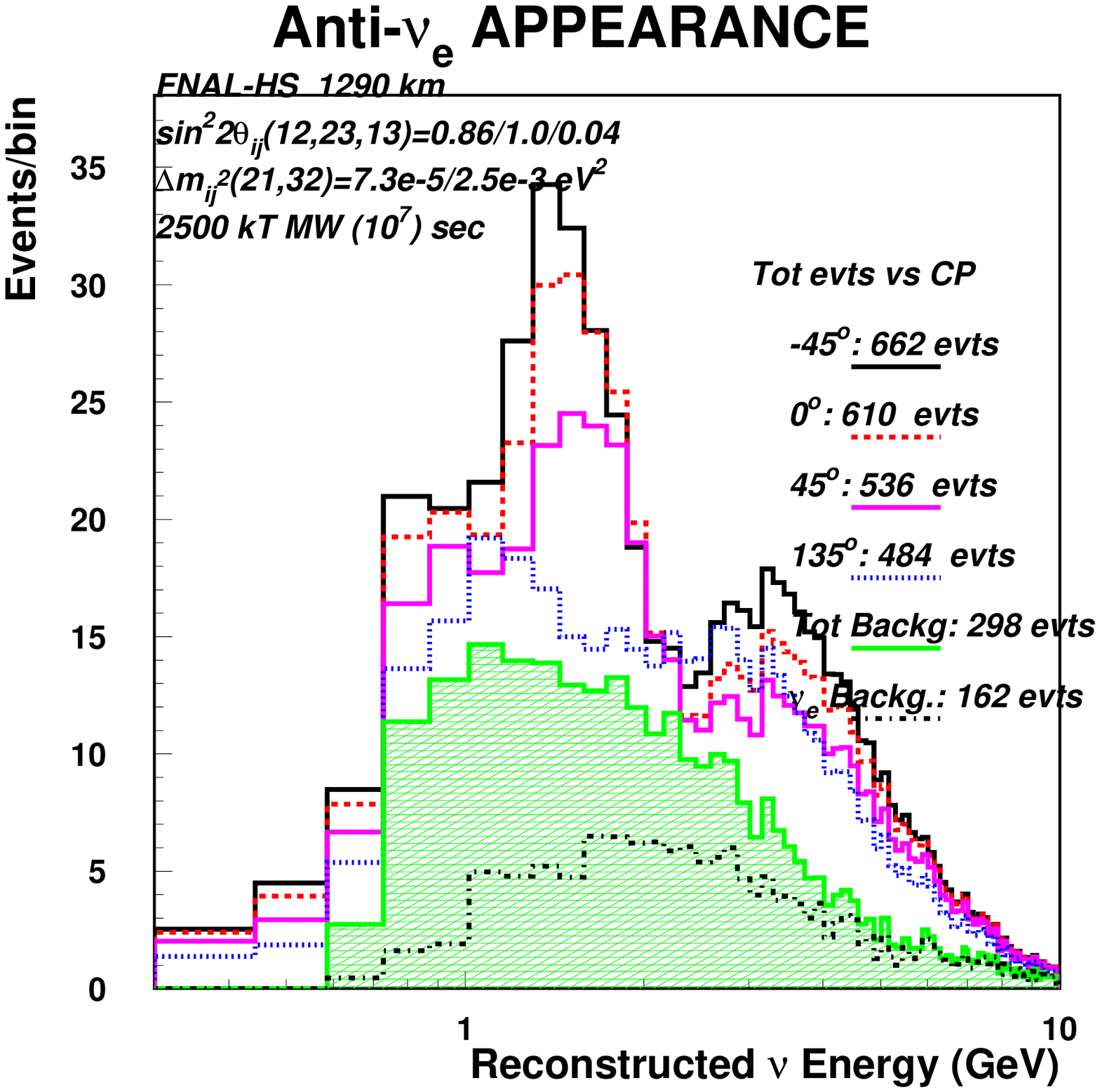}
  \caption[Spectra for $Anti-\nu_e$ appearance]{
Charged current $\bar\nu_e$ interaction spectrum after cuts to reduce background.
The oscillation parameters assumed are indicated in the figures. We have used the value of $\sin^2 2 \theta_{13} =0.04$ for illustrative purposes.  
The CP phase is changed for the various curves as indicated in the figures. 
For these figure  the mass hierarchy is assumed to be "normal". 
These spectra assume the same amount of anti-neutrino running with the 
same intensity proton  beam as in Fig. \ref{app}. 
The spectra  are shown for 2540 (top)and 1290 (bottom) km
distances.
}
   \label{antiapp}
  \end{center}
\end{figure}

We have simulated the results of exposures to neutrino and anti-neutrino
beams from the two laboratories assuming a value of 
$\rho_{Earth}= 3.4 gm/cm^3$, to
take into account the matter effect of neutrinos traversing the long distances
(1300 km from Fermilab, 2540 km from BNL) from those sources to
the Homestake detector. 
The difference between assuming a constant matter density or taking account of 
the density variation is small for these distances through the Earth.
We have also chosen 0.04 as the value of 
$\sin^2 2 \theta_{13}$ in the simulation because a larger value
is likely to be measured in the next few years; a smaller value will
be pursued by this program. Indeed, the CP-test remains viable in the 
Homestake detector even if $\sin^2 2 \theta_{13}$ is as small
as $\stackrel{<}{\sim}$ 0.005. 
At that level,  interest in part shifts to the question of why the 
"solar" and "atmospheric" mixing parameters are so large, and yet the
third mixing angle, which must exist in a 3-generation system,  
is so small. The neutrino 
program would take this as a new challenge, not  an obstacle, 
but  a clear indication that physics beyond the oscillation description 
underlies
the present experimental observations. It is  for this reason among
others that high precision studies of neutrino oscillations are listed  
as a goal 
of the Homestake long baseline neutrino program.
The results of the simulation are shown in Figs. \ref{disap} 
through \ref{antiapp}.

Fig. \ref{disap} shows the nodal structure that will be seen in 
the spectrum of $\nu_\mu$ charged current reactions. 
The same nodal structure is expected in anti-neutrino spectrum.
%identity of the structures of the $\nu_{\mu}$ and anti-$\nu_{\mu}$ 
%disappearance plots as a function of $\nu_{\mu} (\bar{\nu}_{\mu})$ energy. 
Since the anti-neutrino data would be products of a completely separate
accelerator run from the run for neutrino data, a  comparison
of the $\nu_{\mu}$ and $\bar{\nu}_{\mu}$ energy structures would establish
a common energy scale with a given empirically determined error for 
the two energy distributions.

Confirmation of the energy scale (or equivalently $\Delta m^2_{31}$ scale) 
would come from
comparison of the $\nu_{\mu}$ and $\nu_e$ energy distributions in
Fig. \ref{disap}, Fig. \ref{app}, and Fig. \ref{antiapp}, 
which show the locations of 
the $\nu_e$ and $\bar\nu_e$ peaks in the valleys between 
the $\nu_{\mu}$ and $\bar\nu_\mu$ survival  peaks. 
A close examination of Fig. \ref{app} and Fig. \ref{antiapp} shows 
similar spectral features 
except that in the case of normal (reversed) hierarchy the anti-neutrino 
appearance spectrum will be 
suppressed (enhanced) at higher energies.  The effect is dramatic for 
both 1300 and 2540 km distances. 

A large fraction of the CP 
information is gained simply by identifying the nodal structures 
at  well calibrated locations in energy. The position of the 
peaks in energy and their amplitude will give us an 
internally consistent picture of the $\nu_\mu \to \nu_e$ appearance
observation. A detailed fit would yield values 
for the three parameters, $\theta_{13}$, $\delta_{CP}$, and mass hierarchy,
in the sensitive region as shown in Fig. \ref{fig2}. 
Any inconsistencies in the fit could indicate presence of new physics
beyond the 3-generation neutrino model.  

%We emphasize that the
%comparisons above do not depend on $\delta_{CP}$ in any way; their
%purpose would be primarily to normalize the energy scales of the data
%taken in independent neutrino and anti-neutrino exposures: and secondarily,
%to test the internal consistency of the data as a whole.
%The first part of the last statement above does not apply to a comparison
%of the $\nu_e$ and $\bar{\nu}_e$ energy distributions, shown in Fig. 7
%(reproduction of Fig. 4) and Fig. 8 (reproduction of Fig. 6). In this
%comparison, both the matter effect and a non-zero value of $\delta_{CP}$
%are expected to contribute differently to Figs. 7 and 8, the former in the 
%energy region ..., and the latter in the region..., depending on the value of 
%$\delta_{CP}$. Moreover, the amplitude and energy dependence of the matter 
%effect are well known and can be accurately calculated; consequently,
%the matter effect can be factored out of the distributions in 
%Fig. 7 and 8 with 
%small uncertainty, leaving any remaining 
%difference to be ascribed to a non-zero 
%value of $\delta_{CP}$ or, possibly, new physics not included in the
%oscillations formalism.

The capability to predict precisely the energy distributions resulting
from the two data-taking runs described here depends, first, on knowledge of
the measured values of the oscillation
 parameters and their errors, $\Delta m^2_{ij}$
and $\sin^2 2 (\theta_{ij}$), which will be improved in this program. Second,
that capability will depend on the 
ability of making calculations of the event yields and spectra with 
small errors. 
Apart from the tests for consistency
described above, we will also rely on the positive experience obtained
in previous experiments with massive neutrino detectors 
(Kamioka-II, SK, and MINOS\cite{minos})
that have demonstrated good agreement between observation and calculation.

%with small error if the experimental constants involved, e.g., amplitudes
%and energy distributions of the $\nu_{\mu}$, $\nu_e$, $\bar{\nu}_{\mu}$ and
%$\bar{\nu}_e$ fluxes, etc. are known.

The analysis envisioned  above---with its several checks against error
and ambiguities---provides a particularly sensitive search for 
a non-zero value of $\delta_{CP}$
because it uses both the energy dependence of the CP effect as well as the 
neutrino/anti-neutrino asymmetry. 
A less sensitive analysis is possible using neutrino data alone,
 in which the parameter 
$\delta_{CP}$ is measured  explicitly for $\nu_e$ oscillations along
with $\sin^2 2 \theta_{13}$ using the energy dependence alone. 
 The virtue of this mode of analysis is that
it does not require antineutrino data and therefore can be 
carried out when the
neutrino exposure is completed---two to three years after neutrino 
data-taking begins.

\section{Nucleon Decay}

\begin{figure}[htbp]
  \begin{center}
 \includegraphics*[width=0.5\textwidth]{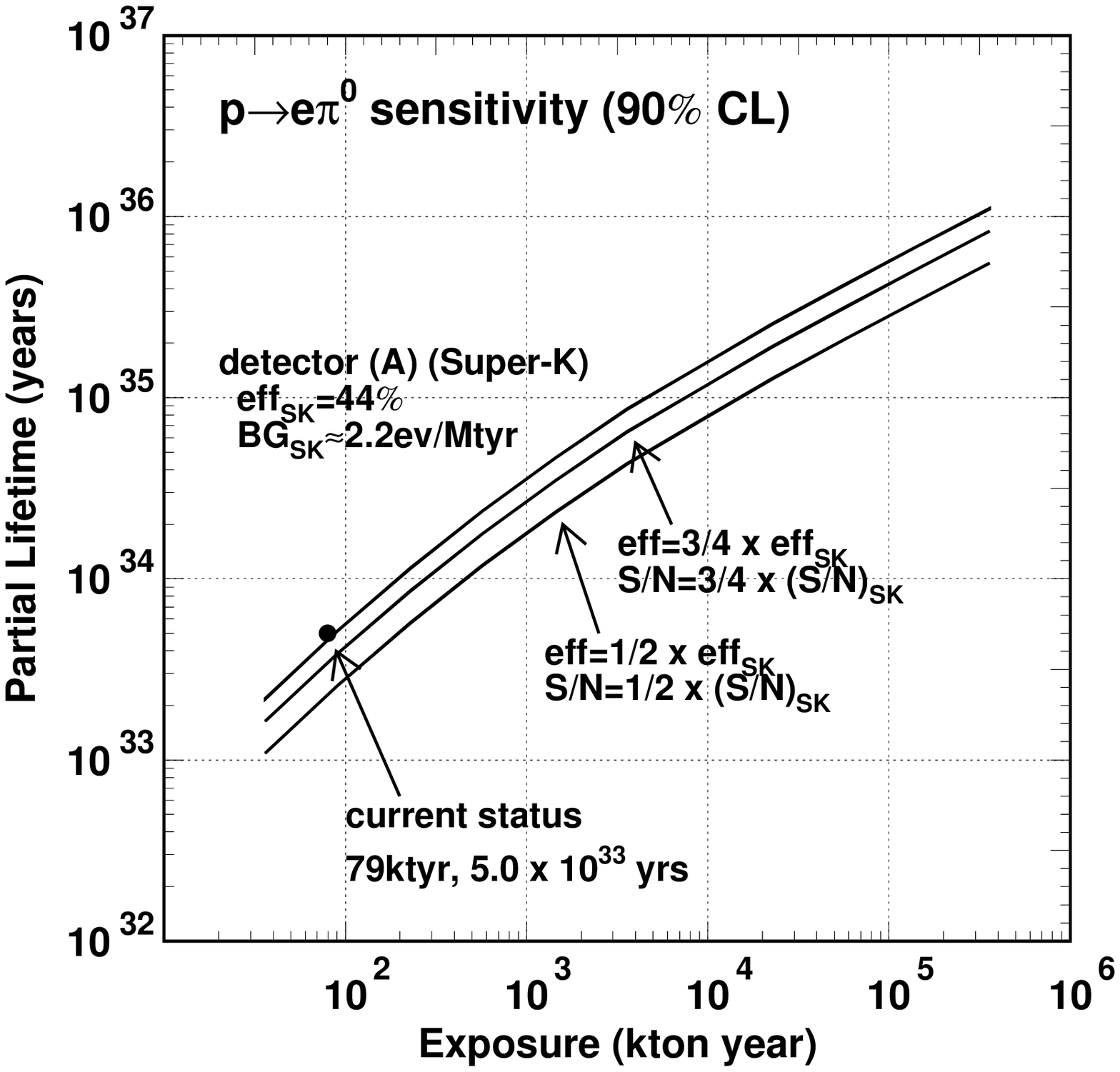}
  \includegraphics*[width=0.5\textwidth]{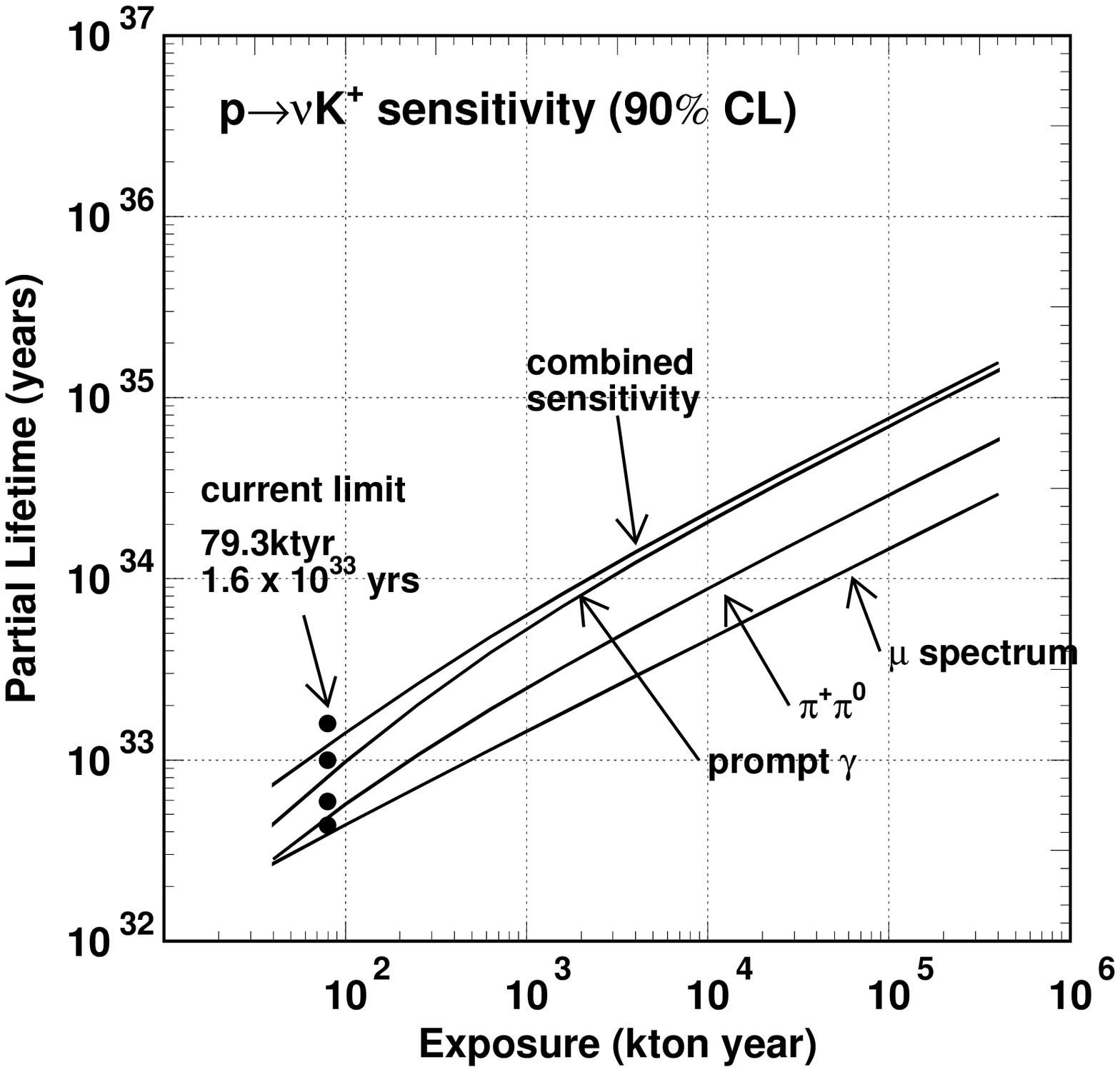}
  \caption[Proton decay sensitivity]{
Projections of proton decay sensitivity for a water Cherenkov detector
from Shiozawa\cite{shiozawa}.  Top plot shows the sensitivity
for $p\to e^+ \pi^0$; the three curves are for three different assumptions 
for the effciency and background reduction in a new detector 
compared to Super-Kamiokande. Bottom plot shows the sensitivity for 
$p\to K^+ \bar\nu$; the plot assumes that there will be some 
improvements in the reconstruction of these events for larger exposures.  
}
   \label{pdk}
  \end{center}
\end{figure}

While current experiments  show that the proton lifetime exceeds about 
$10^{33}$ years, its ultimate stability has 
been questioned since the early 1970s in the
 context of theoretical attempts to arrive at 
a unified picture of the fundamental particles, the quarks and leptons 
 and of their three forces: the strong, 
electromagnetic and weak. These attempts of 
unification, commonly referred to as Grand Unification,
 have turned out to be supported empirically by 
the dramatic meeting of the strengths of the three forces,
 that is found to occur at high energies in the context of so-called 
Supersymmetry, as well as by the magnitude of neutrino 
masses that is suggested by the discovery of atmospheric
 and solar neutrino oscillations. One of the most crucial 
and generic predictions of grand unification, however, 
is that the proton must ultimately decay into leptonic 
matter such as a positron and a meson, revealing quark-lepton unity.
 A class of well-motivated theories of grand unification,
 based on the symmetry of SO(10) and Supersymmetry \cite{babu:1998wi}, which
 have the virtue that they successfully describe the masses
 and mixings of all quarks and leptons including neutrinos,
 and which also explain the origin of the excess of matter 
over anti-matter through a process called leptogenesis,
 provide a conservative (theoretical) upper limit on
 the proton lifetime which is within a factor of ten
 of the current lower limit. This makes the discovery 
potential for proton decay in the next-generation  high. 
At present, the best limit on the classical mode, 
 $p \to e^+ \pi^0$  ( $> 5.4 \times 10^{33}$ yr, at 90\% CL) 
comes from a 92 kton-yr exposure of Super-Kamiokande\cite{skpdk}.
 The detection efficiency of 44\% is dominated by final-state 
$\pi^0$ absorption or charge-exchange in the nucleus, 
and the expected background is 2.2 events/Mton-yr. 
The mode $p\to K^+ + \bar\nu$, is experimentally 
more difficult in water Cherenkov detectors due to 
the unobservable neutrino. The present limit from 
Super-Kamiokande is the result of combining several 
channels, the most sensitive of which is $K^+ \to \mu^+ \nu$ 
 accompanied by a de-excitation signature from the 
remnant $~^{15}N$ nucleus. Monte Carlo 
studies suggest that this mode should remain 
background free for the foreseeable future. 
The present limit on this mode is  $> 2.2\times 10^{33}$ yr (90\% CL). 
Since the next step in nucleon decay very likely
 needs an order of magnitude improvement, 
continued progress inevitably requires much 
larger detectors.  Out of the various technologies discussed 
 water Cherenkov appears to be the only one capable 
of reaching lifetimes of $10^{35}$ years or greater. 
In Fig. \ref{pdk}  we reproduce projections
for proton decay sensitivity for the above discussed modes
 for larger water Cherenkov detector 
exposures\cite{shiozawa, uno}.

The first advantage of a water Cherenkov detector is that 
it is a proven technology that has been perfected 
over several decades. The second advantage is often overlooked:
it is perhaps the only detector with the capability of having a
vast dynamic range in detection of 5 MeV to 50 GeV. This makes it a 
unique instrument ideally suited  to take advantage of a
deep shielded location.  The dynamic range is very important for 
detection of various proton decay modes as well as astrophysical 
neutrinos.

 Water Cherenkov detectors 
are in operation in Japan (Super Kamiokande 
with a total mass of ~50 kT)\cite{sknim} and in Canada 
(the Sudbury Neutrino Observatory, SNO, with 1 kT of
 $D_2O$ and 5 kT of $H_2O$)\cite{snonim}. 
Most of the decay modes that were searched
 for in the first generation detectors required 
only modest depth. IMB operated successfully at 
a depth of 2000 feet \cite{imbnim}. The proposed 
depth here  would reduce 
the muon background by about a factor of 20 
with respect to Super-K and certainly help 
in the observation of modes with a low 
energy component or those influenced by fast neutron background.

\section{Summary of Construction Planning, Budget and
 Timetable for the Caverns 
 in the Homestake Mine \label{summ_const}}

 Plans for the construction of chambers for a multiple module megaton Cherenkov
detector at the Homestake Mine have gone through a number 
of evaluation and design stages consisting of 
rock strength and stability evaluation, chamber design and
 layout, construction planning,  sequencing, and development
 of budget and timetable. Here is a summary of these steps.

\subsection{Characteristics of Homestake Rock, Stability of Large, Deep Caverns
and Coring Plans for the Deep Laboratory Area}

\begin{figure}[htbp]
  \begin{center}
 \includegraphics*[width=\textwidth]{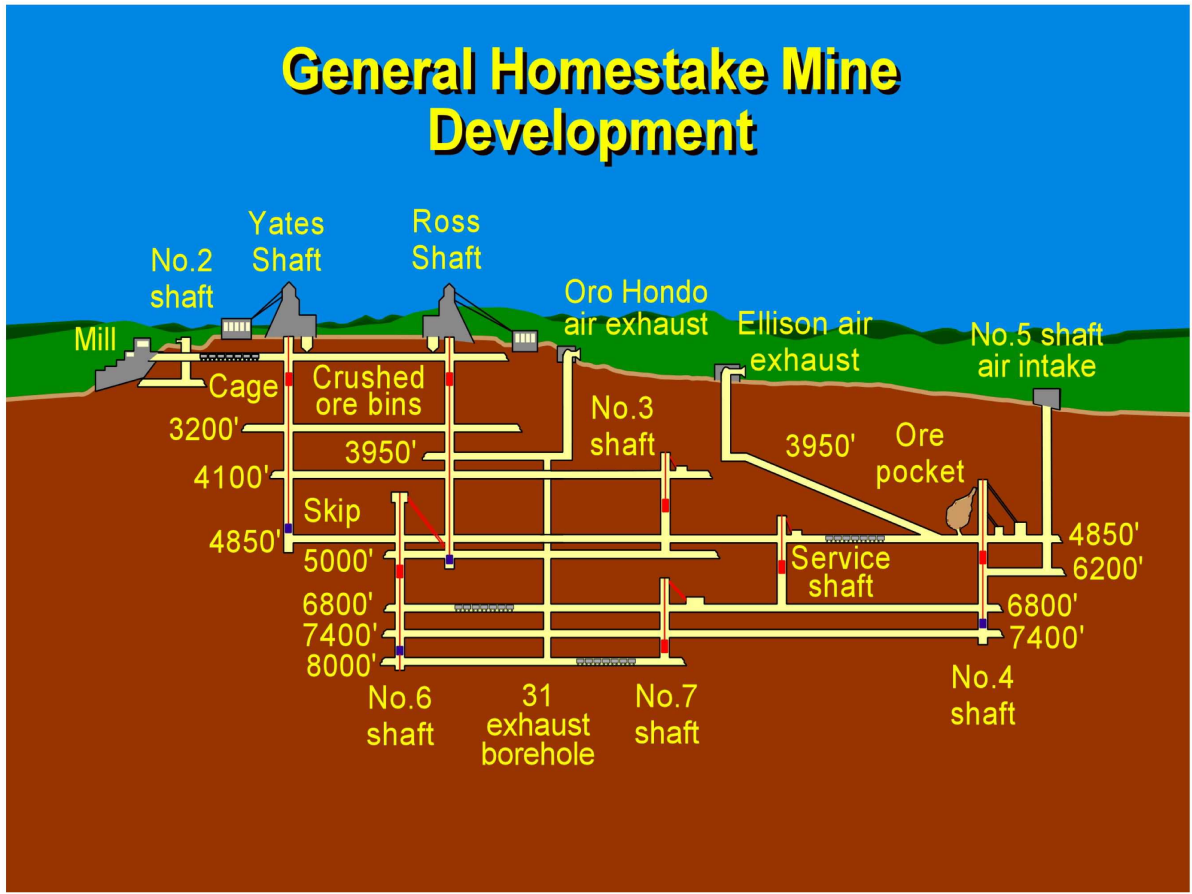}
  \caption[Homestake mine schematic]{
Schematic diagram of the Homestake mine.  

}
   \label{hs1}
  \end{center}
\end{figure}

With the exception of several years during World War II, the Homestake
 Gold Mine in Lead, South Dakota has been in continuous operation from
 1876 until it ceased active operation in 2001\cite{homestake}. During 
that period, over 40 million ounces of gold have been produced. This
 production involved the excavation of several hundred million tons
 of ore from the underground mine as well as over 170 million tons of
 excavated ore and waste rock from the Open Cut (pit) surface mine.
The Homestake Mine now involves over 160 miles of underground tunnels on
 numerous levels down to 8000 ft. below the surface, as shown in 
Fig. \ref{hs1}.

%\begin{figure}[htbp]
%  \begin{center}
% \includegraphics*[width=\textwidth]{1-homestakexsection.eps}
%  \caption[Homestake mine cross section]{
%Cross sectional cut of the Homestake mine development. 
%}
%   \label{hs2}
%  \end{center}
%\end{figure}

The Homestake Mine geology has been extensively studied (see the U.S.
 Geological Survey Bulletin 1857-J (1991) 
and the references cited therein).
 The strength characteristics of the rock have
 been thoroughly studied and
 measured. The advantage of locating detectors deep underground is shown 
quantitatively in Fig. \ref{fig1}.
The Homestake Mining Company (HMC) has constructed several large,
 deep underground chambers. Among these are an equipment repair shop at
 the 7400 ft. level and an air conditioning plant at the 6950 ft. level.
These excavations at great depths provide a strong
 indication that large excavations at depths of 4850 ft. to 7000 ft. 
can be constructed and will remain stable for  multi-decade periods. 
In addition, the construction of 
an array of very large chambers provides a unique opportunity for
 the rock mechanics and geophysics communities to monitor the behavior
 of such structures for long periods.
%Fig. \ref{fh} is an example of such modelling in which 
%the stress in the rock mass between the cavities is calculated. 
This can be monitored by placing sensors in the  drifts that connect 
the detector stations\cite{fairhurst}.  
 This would involve pre-excavation
 computer modeling of the structure, the installation of strain and stress
 gauges during construction and the post-excavation monitoring of these
 instruments. The backfilling of excavations in normal mining precludes
 long term monitoring and the irregular shapes of ore deposits limits the
 geometrical range of the excavations that can be studied.

%\begin{figure}[htbp]
%  \begin{center}
% \includegraphics*[width=\textwidth]{figs/flac3d-fh.eps}
%  \caption[FLAC3D calculation]{
%Example of sress distribution with maximum horizontal principle
%stress directed parallel to diameter.
% Stress values and rock strengths were chosen arbitrarily
% to illustrate that drifts excavated 
%(before cavity construction) will be subjected to various stresses 
%at different distances from the caverns. This variation
% can be used in developing understanding of how rock 
%responds to changing stress conditions. 
%\cite{fairhurst}  
%}
%   \label{fh}
%  \end{center}
%\end{figure}

The region of monolithic rock being
 considered for the deep underground Cherenkov
 detector has not been mined or explored in detail. 
By extrapolating from above
 and from the west side, this rock region appears homogeneous, and with
 few intrusions. In this document 
we are proposing that the initial 3 chambers be located on the 4850 ft level 
close to the Yates shaft and the Chlorine experiment. The rock in this 
region is better known than the rest of the Yates formation, and therefore 
the confidence in the cost and time estimates are higher. On the other hand,
before beginning any excavations, it would be advisable
 to carry out extensive 
coring studies of this volume of rock. Cores can be drilled
 down from tunnels on the 4850 ft. level. These cores 
would verify the rock strength 
characteristics and would provide information about the direction of the 
rock  grain.

\subsection{Status of the Homestake Mine}

Earlier this year, 2006, title to and possession of the Homestake Mine
was transferred from the Barrick Gold Corporation to the State of
South Dakota.  The State has designated the South Dakota Science and
Technology Authority to operate the Mine as a Deep Underground
Laboratory and has provided the Authority with \$35 million in State
funds and \$10 million in Federal funds for this purpose.  The
Authority anticipates that it will spend the rest of 2006 on various
underground remedial activities and to reestablish laboratory
facilities, power distribution, ventilation and communications system.
It anticipates opening the Underground Laboratory for experiments in
late 2007.  The initial laboratory area will be at the 4850 ft level,
4200 meters water equivalent deep. 
 The Authority anticipates that its existing funding is
sufficient to provide basic Laboratory operations through 2012.

In addition to the above funds, Mr. Denny Sanford, the CEO of Premier Bank and
Premier Bankcard, has offered to donate \$70 million to the Homestake
Laboratory, giving the Lab a total of \$116 million in non-NSF funding.

The last indication was that the water in the lower reaches of the
mine has reached the 6200 ft level, about 1350 ft below the 4850 ft
laboratory level.  Since about 2/3 of the water inflow is surface
water that enters above the 4850 ft level, one of the 
remedial activities in 2006 is to intercept this water and pump it
back to the surface using the existing mine pumping system.  Once NSF
funds are made available, the Authority will pump out the water from
the deeper regions of the mine and establish a second, deeper
laboratory region at the 7400 ft level.

As noted in Fig.  \ref{linout}, the initial three 100 kiloton fiducial mass
 chambers are to
be constructed adjacent to the site of the Chlorine Solar Neutrino
Detector.  Each of these chambers will be about 500 ft away from that
site.  The rock here is well known.  The Chlorine Detector chamber was
excavated in 1965 and has been completely stable since then.  In
addition, the Yates rock dump and rock hoist system is only 500 ft
from the Chlorine chamber so that waste rock from these first three
chambers need only be transported about 1000 ft to the rock hoist
system.  The existing Chlorine Detector laboratory area may also serve
as an initial and temporary staging area for this three chamber
complex.

\subsection{Determination of Excavation Stability}

A preliminary 2 dimensional large chamber stability evaluation was
carried out in the fall of 2000 by members of the Rock Stability Group
at the Spokane Research Laboratory of NIOSH (National Institute of
Occupational Safety and Health). This evaluation indicated that stable
chambers with dimensions in excess of 50 meters could be constructed
at depths of 7000 ft or more at the Homestake Mine.  In the fall of
2001, rock samples from the Yates formation were taken to the Spokane
Laboratory and strength and stress analyzed.  These measurements
provided more specific input for a three dimensional stability
analysis of large excavations as a function of depth in the Yates rock
formation in the Homestake Mine. This 3D analysis involved a finite
difference evaluation using the FLAC3D program. These results
\cite{stab1}
 were compared with the empirical prediction charts of \cite{charts}.
 The conclusions were that 50 meter diameter by 50 meter high chambers
 could
 be safely excavated and would be stable for long term occupancy 
at depths up to 2150
 meters and probably somewhat deeper.

The results of this evaluation were presented in a paper at the Fifth
 North American Rock Mechanics Symposium in July 2002 \cite{stab1}.
  For the measured Yates formation rock parameters (Tunneling Quality Index), 
the Grimstad and Barton specifications are 4 - 5 cm of unreinforced
 shotcrete applied to the rock surface.  We have decided to install a 50 cm 
thick concrete shell on the vertical cylindrical inside of the excavation.
  This is equivalent to a two order of magnitude safety factor using the
 Barton and Grimstad parameters \cite{barton}.

\begin{figure}[htbp]
  \begin{center}
 \includegraphics*[width=\textwidth]{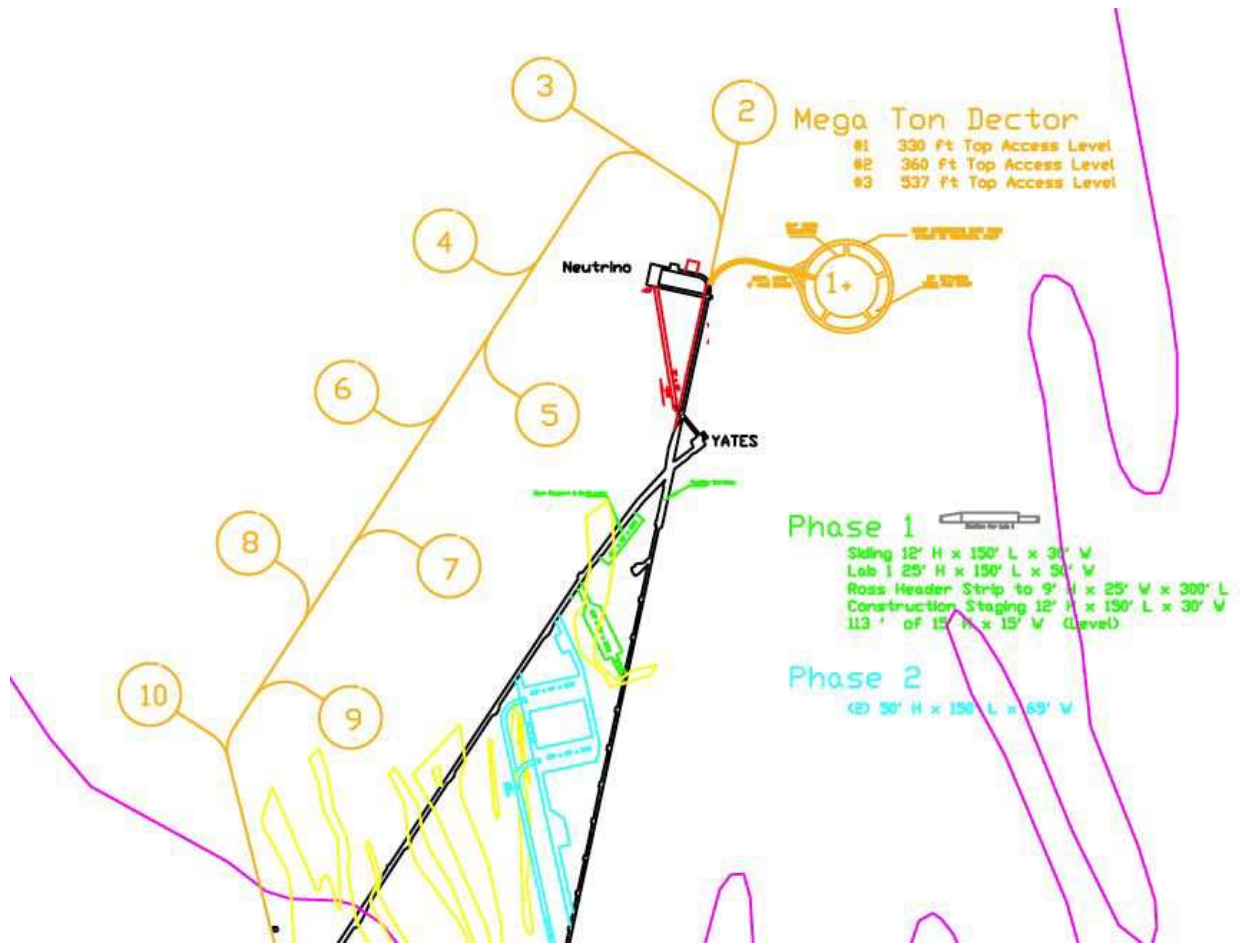}
  \caption[Linear layout of 10 Caverns]{
Possible linear layout of 10 caverns at Homestake.
This layout is developed for the 4850 ft level. 
The location of the Davis Chlorine cavern is indicated by 
the word ``neutrino''. 
}
   \label{linout}
  \end{center}
\end{figure}

The FLAC3D analysis also indicated that in order to isolate the stress
 fields of one chamber from the adjacent one, these chambers should be 
separated by at least 3 chamber diameters.  
%That specification then
% determined the radius of the circle  on which the chambers will reside. 
% Finally, the circular layout was adjusted to avoid known inhomogeneities
% in the rock. See Fig. \ref{layout}.
These conclusions were confirmed by the report of the NSF site selection
committee 
\cite{siteselection}.
Recently a linear layout has been envisioned for the 4850 ft level. 
Such a scheme is shown in Fig. \ref{linout}.   This scheme has many advantages
mostly due to the  proximity of the 
caverns to the well developed area between the
 Yates and the Ross  shafts.

%\begin{figure}[htbp]
%  \begin{center}
% \includegraphics*[width=\textwidth]{figs/layout.eps}
%  \caption[Layout of 10 Caverns]{
%Possible semicircular layout of 10 caverns at Homestake.
%This layout was originally developed for the 6950 ft level, but it
%could be easily adapted to 4850 ft.  
%}
%   \label{layout}
%  \end{center}
%\end{figure}

\subsection{Construction of Multiple 100 kiloton Chambers
 in the Homestake Mine}
\label{multiple}

Using the results of the stability evaluation, a group of former Homestake mining 
engineers, (Mark Laurenti--former Chief Mine Engineer, Mike Stahl--former
 Mine
Production Engineer and John Marks--former Chief Ventilation Engineer)
 designed
an array of ten 100 kiloton water Cherenkov chambers. The criteria used 
in this design were, a minimum of 50 year safe occupancy of the chambers,
 independent ventilation and access system for each chamber so that completed
 chambers can be used for research while additional chambers are under
 construction, and a structure that will permit a plastic lined water 
tight and radon reducing structural enclosure.  

\begin{figure}[htbp]
  \begin{center}
 \includegraphics*[width=\textwidth]{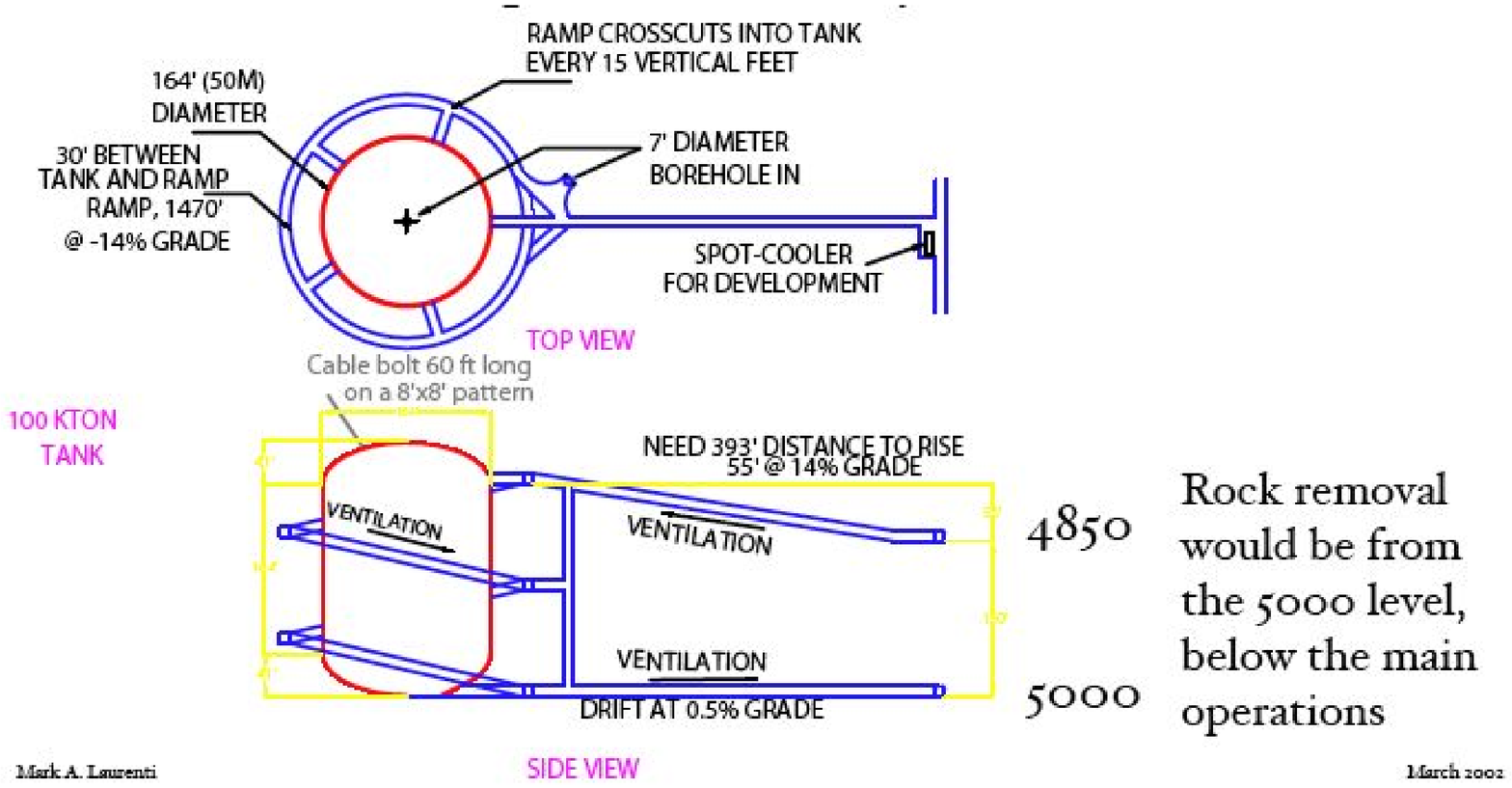}
  \caption[Details of a single cavern]{
Details of the construction of a single cavern. 
}
   \label{detail1}
  \end{center}
\end{figure}

%The original detector construction plan was for a detector
% array at the 6950 ft depth.  We have adapted that layout for
% the 4850 ft level. 
 The design involves a detailed construction plan, a rock reinforcement 
plan with cable and rock bolts and a cylindrical concrete liner,
 and a coordinated water handling, ventilation and chamber access plan. 
 The top of each chamber is
 connected to the 4850 ft. level via a horizontal, radial 
tunnel. A similar tunnel connects the bottom of each chamber to a
 tunnel at the 5000 ft. level.
Fresh air will be sent to each chamber via the top tunnel and 
exhaust air removed via the bottom tunnel, thus providing independent air
 supplies to each chamber. Plan and elevation views of  a single 
cavern
  are shown in Fig. \ref{detail1}. 
%in Figs.~\ref{f:array.jpg} and ~\ref{f:module.jpg}.
The chamber dimensions are nominal 
>and will be modified as cores and other rock information are acquired.

\begin{figure}[htbp]
  \begin{center}
 \includegraphics*[width=\textwidth]{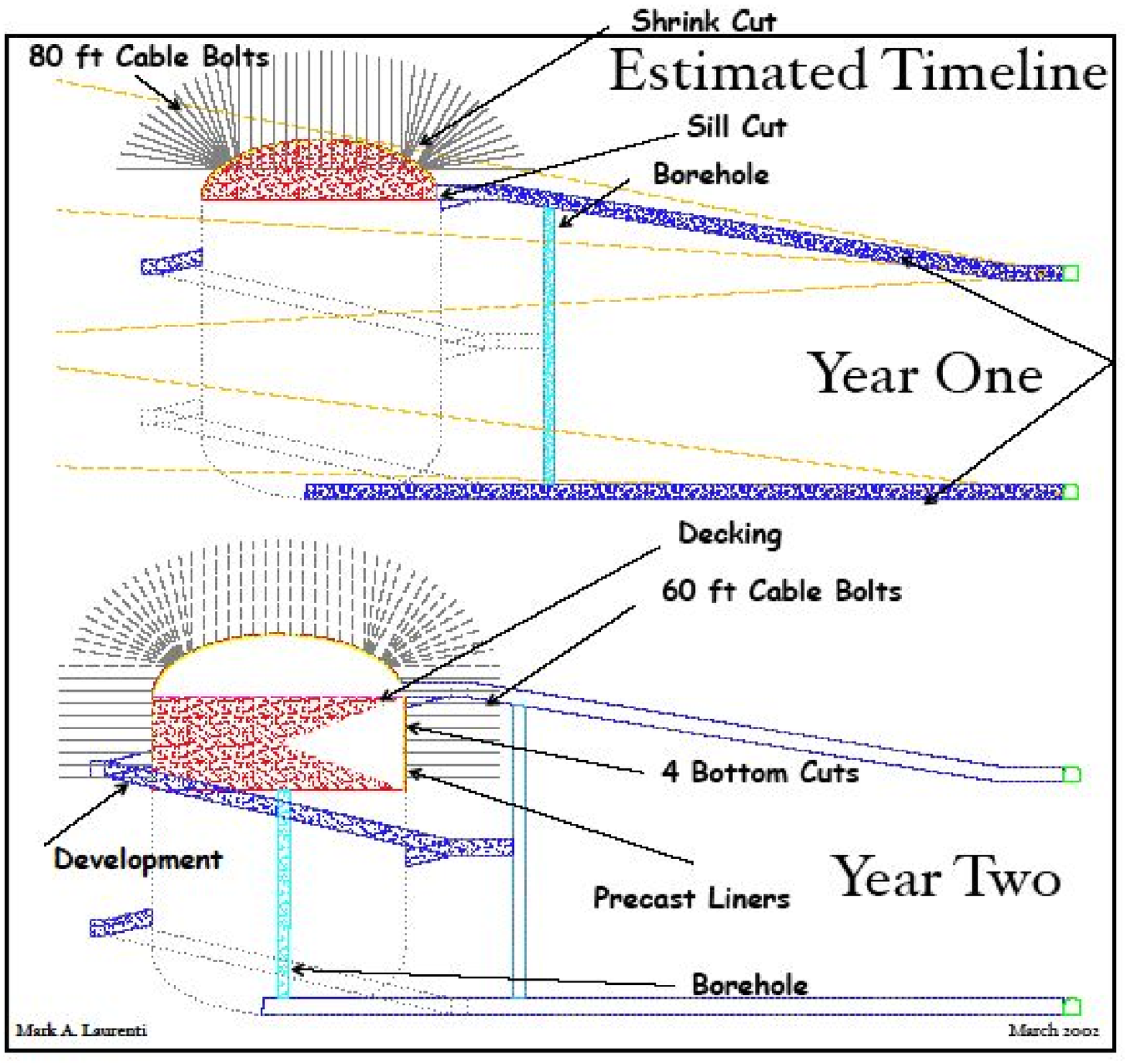}
  \caption[Cavern construction year 1 and 2]{
Details of the construction of a single cavern. 
}
   \label{yr12}
  \end{center}
\end{figure}

\begin{figure}[htbp]
  \begin{center}
 \includegraphics*[width=\textwidth]{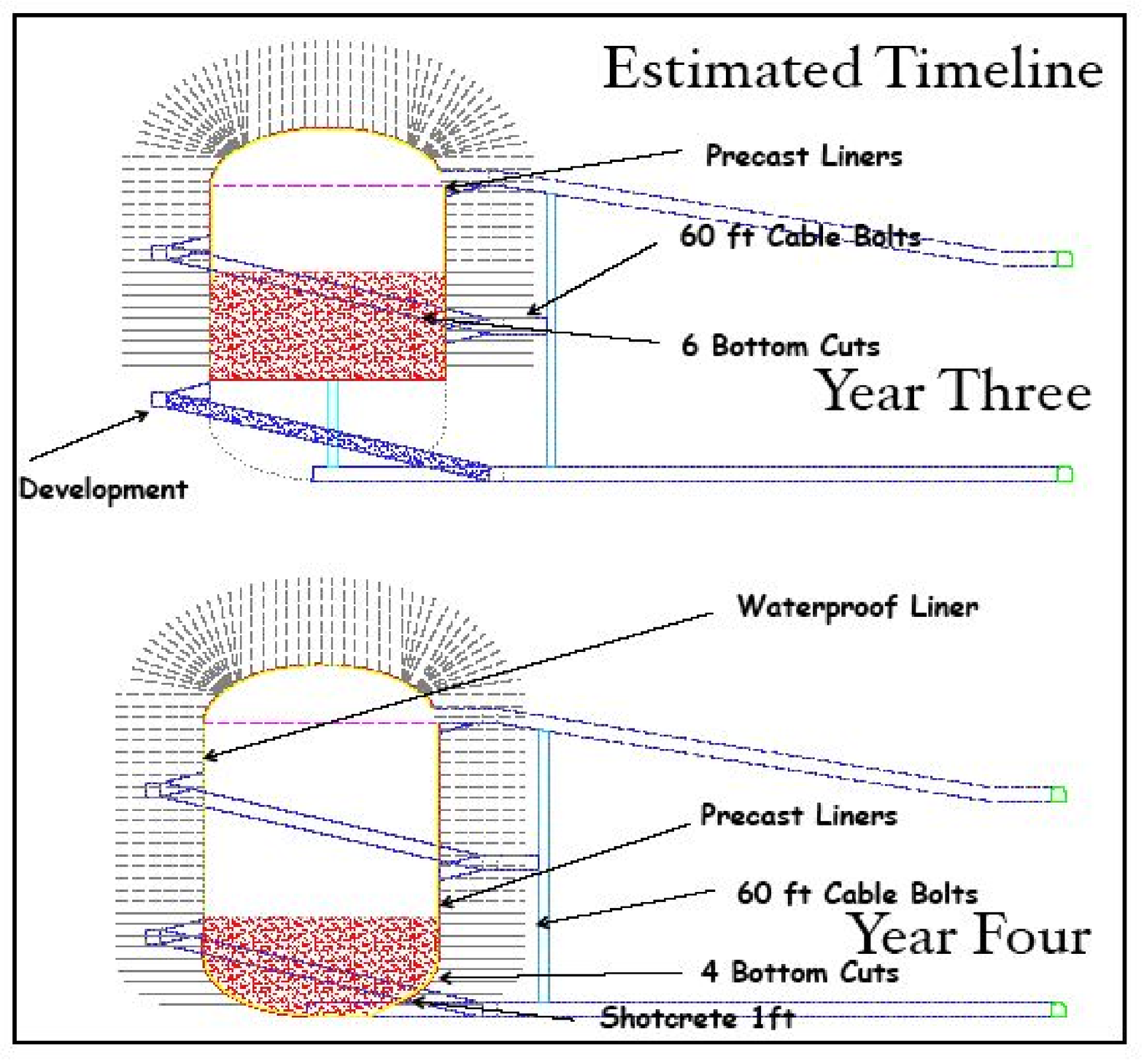}
  \caption[Cavern construction year 3 and 4]{
Details of the construction of a single cavern in year 3 and 4.  
}
   \label{yr34}
  \end{center}
\end{figure}

In Fig. \ref{yr12} and Fig. \ref{yr34} we show the sequence of steps involved
over a four year period in the excavation of one chamber. Our plan is to 
excavate a group of three chambers at once. 
%of 
%the tank excavation over a 4 year period. Below we will discuss the 
%details.  
%A single radial tunnel will access a group of three chambers. 
% This tunnel is directed
% toward the central chamber of the group with side branching 
%tunnels to the other two
% chambers of the group.  The branching locations are staggered to
% avoid a three tunnel
% meeting point.   
An equipment maintenance room will be
 constructed alongside the top access tunnel.  After construction
 of the three chamber
 group is completed, this room will be converted into an electronics
 and water purification
 facility that serves this three chamber complex.

Two 2.4 meter diameter vertical bore holes will be constructed between
 the upper and lower
 tunnels.  One of these bore holes will be outside the chamber and
 the second in the
 center of the chamber.  The bore holes serve three purposes.  One,
 they complete the air
 flow path between the fresh air input of the upper tunnel and the exhaust
 of the lower
 tunnel.  Two, they provide a drop path for excavated rock to the lower
 tunnel for removal.
  Three, the bore hole in the center of the chamber provides
 an expansion region for
 blasting of rock.  By blasting rock inward from an ever
 expanding set of concentric 
circles, we will avoid fracture of the outer rock walls that will
 serve as the chamber
 containing walls.

Removing waste rock and associated rock dust via the bottom air
 exhaust tunnel, 
will prevent rock dust from one chamber contaminating the fresh 
air supply of another
 chamber. Once construction is completed, the bottom chamber to
 tunnel connection will
 be sealed.

Once the access tunnels and bore holes are completed, construction
 of the dished dome of
 the chamber will begin.  The first step is to excavate a 53 meter
 diameter by 3 meter
 high section.  80 ft. long vertical cable bolts will be installed
 into the roof of this
 excavation as that work proceeds. Once this section is completed, 
the excavation will
 be extended upward  to form the dome. As this domed excavation
 progresses, some of the
 80 ft. cable bolts will be cut.  When completed, the dome will have 
 60 ft. cable bolts
 in the center and longer cable bolts at larger radii with 80 ft.
 cable bolts along the
 outer edge of the dome.  The stress diagrams in the diagrams in 
calculations of Tesarik, Johnson, and Zipf 
\cite{stab1}  
 indicate that
 the regions of greatest stress are along the outer region
 of the dome, where the cable
 bolts are the longest.

Once the central chamber access tunnels and boreholes are completed,
 work on the two
 side chamber access tunnels and bore holes can begin.  This work
 will be done in parallel
 with the construction of the dome of the central chamber.  Thus,
 once the central chamber dome is finished, about one year 
after beginning of construction,
 we will be ready to
 begin the domes on the two side chambers. This parallel construction
 mode with the side
 chambers several months behind the central chamber allows efficient
 use of equipment and
 personnel while at the same time permitting us to test the 
construction plan on the
 central chamber before employing it on the two side chambers. 
 This overall construction
 procedure will result in one completed chamber in four years
 with the two side chambers 
available a few months later.

In year two, we will begin on the main, vertical cylindrical
 section of the central chamber. 
 There will be a series of 3 meter high cuts of a 53 meter 
diameter section, essentially the
 same size as the initial cut.  These cuts will be made much
 faster since there is no need for
 vertical cable bolting.  Each of these cuts consists of three steps.

The first step is to drill downward holes for explosives and then to set off these
 explosives.  By using a concentric circular layout and by staggering the
 firing of these
 explosives, the rock will be blasted toward the open central borehole region and
 minimize outward shock waves.  In addition, we plan to drill a series of
 vertical holes
 along the cylinder boundary to prevent outward transmission of shock waves
 and resultant
 fracturing of the cylinder walls.

The second step is to move the blasted rock fragments into the two boreholes so
 that they
 can fall to the bottom.  Front end loaders in the bottom tunnel will move
 the broken rock
 to the rock dump near the shaft for hoisting.

The third step is to install horizontal, radial 60 ft long cable bolts into the
 cylindrical 
rock walls and immediately thereafter install the precast concrete liner sections. 
  These liner sections will be bolted into the rock and will dovetail into each
 other. 
 The joints between liner sections will be grouted so that the inner surface is
 smooth.
 Finally, we will spray on a layer of Mine Guard, a polyurethane coating that
 will serve 
as a inhibitor of radon flow from the walls and simultaneously a very smooth and 
low
 friction surface for the water tight plastic liner that will be installed at
 the end.

Each 3 meter high cut will take 10 weeks, 3 weeks for the blast hole drilling,
 3 1/2 weeks
 for the rock removal and another 3 1/2 weeks for cable bolt installation.
  The concrete 
liner installation will overlap the cable bolt and next layer blast hole drilling. 
 By staggering these activities by about 3 weeks between each of the three
 chambers,
 we can utilize three specialized crews, one for each of these tasks, and move
 them 
from one chamber to the next.  In addition, since separate equipment is used for
 each
 of these three tasks, we efficiently utilize the equipment and avoid having it
 stand
 idle for 2/3 of the time.

In order to permit horizontal entrance into the excavation region for each of the
 three
 meter high cuts, we will build a spiral ramp around each chamber.  Each cut will
 consist 
of a horizontal entry into the chamber region from this ramp.  Once a given layer
 is

 completed, the entry cut will be sealed with a concrete plug.  These ramps will
 permit
 machinery to easily move from one chamber to another or to the maintenance area
 and will
 provide a path for moving some of the blasted rock to the outside bore hole. 
 At the end
 of the chamber construction process, these spiral ramps will serve as secondary
 ventilation links between the input and exhaust air systems and provide convenient
 local access to the bottom of each chamber.  For example, we can carry out water
 purification by withdrawing water from the bottom of the detector, pumping it up 
through a pipe in the spiral ramp to a purification system at the top and then returning 
it to the top of the chamber.

The chamber will have an outer circular walkway at the top of the chamber and a 
central gridwork that will be suspended from the upper dome cable bolts.  Both of these
 will be installed at the time the top cylinder cut is made.  The walkway will provide 
access to the upper ends of the side wall photomultiplier arrays.  The top gridwork will
 provide a mounting platform for the top layer of photomultiplier tubes as well as 
access for various calibration equipment.

The final chamber construction task will be the installation of the inner plastic liners.  
The conventional procedure is to use two liners, one a geotextile that serves as a cushion 
and also an inner fused water tight liner  Each of these materials are supplied on rolls.
  We plan to mount the end of each roll against the rock along the upper walkway and then
 gradually lower the roller and so unroll the material along the vertical side of the 
cylinder.  The geotextile layers will be unrolled so that there is a butt joint between
 adjacent layers.  A tape seal will then be applied to these joints.  Access will be via
 a bosun chair suspension from the walkway.  Once the geotextile layer is complete, the
 waterproof layer will be lowered.  In this case, we will use an overlap joint which will
 then be heat sealed, again via the bosun chair suspension.

The plastic liners along the bottom will be extensions of the side liners.  Before being 
brought into the mine, the bottom section of each roll will be cut into a triangular shape
 to conform to the bottom.  We may add an additional bottom plastic overlay to insure that
 this section is truly water tight.  It is worth noting that this multi-layer water tight
 construction is conventionally used to seal the bottoms of reservoirs and other bodies of
 water.

\subsection{Construction Timetable and Cost}
\label{twotimetable}

Mark Laurenti has worked out a detailed timetable and budget for the
construction of these chambers including initial rock evaluation
coring, construction of both top and bottom access tunnels, removal of
waste rock, maintenance of mining equipment, etc.

The excavation process consists of continuous repetition of three
separate tasks (1) drilling and blasting of rock, (2) removal of the
rock rubble, and (3) installation of rock and cable bolts to stabilize
the freshly exposed rock walls. Each excavation cycle is about 10
weeks with 3 weeks for each of the above three steps. There is a cost
savings in excavating three chambers at the same time, with a three
week phase shift between steps in each module. This arrangement
permits each of the three specialized crews to move from one
excavation to the next every three weeks or so and continue using the
same equipment and carry out their specialized tasks. In contrast,
using one crew to sequentially do three different tasks will result in
idle equipment for 2/3 of the time and inefficiency as they switch
from one task to another.

For the three chamber mode, the cost of excavating each chamber in
2002 was \$14.7M per chamber.  We have adjusted that by the standard
inflation index and other factors  to \$22M.  
This includes \$3.6M for the concrete
liner and a 30\% contingency. In contrast, the cost of excavating a
single chamber in 2002 was \$19.1M including liner and contingency
(See Table \ref{laurenti}).  Again, we have adjusted that by the
standard inflation index and other factors 
to \$29.1M.  The total required equipment
cost is the same for both of these construction modes.

Assuming three shifts/day and 5 days/week operation, it will take 208
weeks or 4 years to excavate each 3 chamber group. This time could be
reduced by going to a 6 or 7 day week.  The Homestake Company
frequently operated on a 6 or 7 day per week basis.

\subsection{Rock Removal}

Each 100 kiloton chamber ($10^5 m^3$) will involve the removal of
about 419,600 tons of rock including access tunnels, domed roof,
etc. For three chambers this results in 1,258,800 tons of rock in 4
years or 314,700 tons of rock per year. Since the hoisting capacity of
either the Yates or Ross shaft system is 750,000 tons per year, the
simultaneous construction of three chambers utilizes only 40\% of the
capacity of one of the two existing shaft systems. The Lab oratory has
identified a number of nearby rock disposal sites. Disposal costs per
chamber are expected to be of the order of \$1M, which were included in 
the numbers in the previous section.

\subsection{Equipment Cost}

Since all mining equipment has now been removed, new mining equipment
will have to be purchased or leased. The required equipment, one Face
Drill, two LHD loaders, 2 Bolters, 2 Underground Support Vehicles, 2
Lift Trucks, 1 LH Drill and 2 ITH Drills, costs about \$5 M. It may be
possible to arrange for leases instead of purchasing these
items. Normal equipment maintenance has been included in the
construction cost.  The cost of this equipment should probably be
assigned to this specific task; it will be used full time during the
construction period.

\subsection{Choice of Depth }

There has been considerable discussion of depth necessary for very
large detectors and the costs associated with deep detector
locations. It is clear that the deeper the detector, the lower the
cosmic ray muon and associated particle background. It is always
preferable to have lower background.  The measured cosmic ray muon
flux at the 4850 ft level is 4 muons per square meter per day.  This
corresponds to roughly 0.1 Hz for a single 100 kton detector
module. The beam pulse from FNAL (BNL) will be 10 $\mu sec$ (2.5 $\mu
sec$) long.  This results in less than one cosmic ray muon per year
going through one of the chambers during the "neutrino beam on time"
from FNAL.  This insures that any signal seen during the beam on time
is due to the interaction of an accelerator neutrino.  No special
pattern recognition (to measure event directionality) or veto shield
is required. We anticipate considerable gain in neutrino event
efficiency factors and perhaps in fiducial volume.

\subsection{Summary of Section~\ref{summ_const}}

A detailed construction plan for the construction of three 100 kiloton
chambers in four years at the 4850 ft. depth in the Homestake Mine has
been developed. The total construction cost of the three chambers is
about \$66.1M or\$16.5M/year, as shown in detail in section 4.4. 
This includes 
 a one time purchase of about \$5 M worth of
mechanized mining equipment. The lead time in delivery of the mining
equipment can be used to carry out coring of the rock region in which
the detector array is to be constructed.

In the next section we turn to the details of construction and cost of
one chamber of the array of chambers in the absence of the detector
that fills the chamber. A single chamber 50 m in diam $\times$ 50 m
high is a taxing mining engineering accomplishment. We concentrate on
a single one without detector components, which are discussed later,
because we want to keep separate in our planning the large scale
mining operations---a 50 meter high structure (even without dome) is
the rough equivalent of a twelve story building---which are less
familiar to us.

\section{Design and Cost of One Chamber and
 Projected Cost of
 Multiple Chambers}
\label{designandcost}

Our ultimate goal, the construction of a detector 
with a total mass of a megaton, 
can be best achieved by constructing 10 
separate 50 meter diameter by 50 meter high 
(100 kiloton) cylindrical chambers. 
In this section we consider the plans and the 
cost of constructing one such
 chamber, a 3 chamber system,  and  the 10 
module megachamber system.

\subsection{Single Chamber Construction Costs}
\vspace{-1ex}

A detailed construction timetable and cost for the construction of
these chambers was developed by Mark Laurenti.  It is worth repeating
Laurenti's credentials, the Chief Mine Engineer for Homestake until
the end of 2001, when the Company ceased all underground mining.  The
development of this timetable and budget utilized his experience in
budgeting and planning underground excavation and construction at the
Homestake Mine.  Table~\ref{laurenti} summarizes the 2002 costs for
the construction of a single 100,000 ton chamber.

\begin{table}[ht]
\begin{center}
\caption{\label{laurenti}Summary of Estimated Year 2002 Total Costs (M. Laurenti)}
\begin{tabular}{|l|r|} \hline \hline
Total Labor \& Benefits & \$5,506,656\\ 
\hline Total Equipment Operation & 1,297,210\\ 
\hline Total Supplies (including concreteliner) & 7,760, 490\\ 
\hline Total Contractors & 123,000\\ 
\hline 30\% Contingency & 4, 406, 207\\ 
[1ex]\hline \hline
% & \rule{16mm}{.5mm}\\ 
 Grand Total(2002) & \$19,093564\\ \hline
\end{tabular}
\end{center}
\end{table}

 Adding the non-concrete liner costs and the transport cost to the
 above subtotal gives \$19, 564,560 or \$46.6/ton excavated.  We
 tested the validity of these cost estimates against several other
 measures from the same time period.  One of these involved a cost
 estimate that Dynatec, a highly regarded Canadian mine consulting and
 mine construction firm made in February 2001 as part of a laboratory
 development program. Although Dynatec used a somewhat different cost
 construction, the final Dynatec costs were in reasonable agreement
 with the Laurenti cost estimate.

\subsection{Historic Costs of the Homestake Mining Company}
 
Another way to establish the cost of excavation is to use the historic
cost that the Homestake Mining Company published in its annual
reports. For 1997, HMC gave\$49.88/ton; for 1998, HMC gave
\$34.40/ton; and for 1999, HMC gave \$44.48/ton; an average of
\$42.90/ton. All of these figures are per ton of ore extracted and
processed to extract gold. The costs include all mine operating costs
and exclude only taxes, depreciation and central corporate
charges. Thus, these figures include the cost of gold extraction from
the ore, charges for mine operation such as ventilation, general mine
maintenance, mine operation overhead, etc. The costs also include
charges for exploratory work in looking for additional gold deposits,
the removal of waste rock required to access the gold ore, etc.  A
year in which there was little exploratory work or waste rock removal
would have a low cost per ton while one in which there was significant
such effort would have a higher cost per ton.

Given these considerations, the historic HMC costs are in agreement
with the summary of costs given above, well within the uncertainties
of both numbers.

\subsection{Final Total for  Cost of Construction of the First Chamber.}

To the best of our understanding, the total 2002 cost of \$19,093,564
for the first chamber---including a 30\% contingency---is
well-documented as indicated by Table~\ref{laurenti}.  The detector
components and the funding rate to produce a complete detector module
will be extra and we turn to them in section 6.

\subsection{Costs and Schedules for Multi-Chambers}
\subsubsection{Three Chambers}

We plan to finish three chambers four to five years after the initial
funding becomes available. This can be done either by starting to
excavate three chambers at the same time in the first year, or
somewhat more conservatively by carrying out excavation of a single
chamber during the first year to gain experience, and then proceeding
to the second and third chambers as well as continuing with the first.
 
\begin{table}
\begin{center}
\caption{\label{comparison}Comparison of single 
chamber versus three chamber cost}
\begin{tabular}{|l|l|r|r|} \hline \hline
Estimated Costs (\$MM) & & & \\ \hline \hline
& \# Of Chambers & 1\ \   & 3 \ \ \\ \hline \hline 
Labor \& Benefits & & \$5.51 & \$10.94\\ \hline 
Mining and Construction  & & &\\ \hline 
\ \ \ \ Equipment Operation & & \$1.30 & \$3.89\\ \hline 
\ \ \ \ Supplies & & \$4.51 & \$13.35\\ \hline 
\ \ \ \ Precast Concrete Liner & & \$3.25 & \$9.75\\ \hline 
Other (Outside Contractors) & & \$0.17 & \$0.52\\ \hline 
30\% Contingency & & \$4.40 &  \$11.48\\  [1ex] \hline \hline 
% & & \rule{16mm}{.5mm}  & \rule{16mm}{.5mm}\\ [1ex] 
& TOTAL(2002) & \$19.1 & \$49.93\\ \hline
& TOTAL(2007) & \$29.1  & \$66.1   \\ \hline
\end{tabular}
\end{center}
\end{table}

\subsubsection{Three Chamber Construction Cost}

As pointed out above, the chamber excavation procedure consists of
three approximately equal duration tasks, blast hole drilling and blasting,
blasted rock removal and installation of rock support cables and concrete
liner. Each of these tasks requires specialized equipment and specific work
skills. In the construction of a single chamber, equipment is idle 2/3 of
the time and the mine crew switches from one task to the next
every three weeks or so. Both of these situations lead to inefficiency
in chamber construction and so higher construction costs. There is an
additional a factor, the ratio of supervisory and equipment maintenance
personnel to miners.

By constructing three chambers at once with a phase shift between operations
in each chamber, we can avoid these inefficiencies. Now, we would have
three separate crews, one for each of the above sub-tasks. After the blasting
crew drills the blast holes and blasts the next rock layer in the first
chamber, they would immediately move on to the second chamber and do the
same there. At the same time the rock removal crew would move into
the first chamber. About three weeks later, the blast crew moves to the
third chamber, the rock removal crew to the second chamber and the wall
support crew into the first chamber. Each crew would become expert in
its specific tasks and might be able to reduce the time required to carry 
them out. In either case we anticipate a yearly expenditure rate of
 approximately \$16 - 20M for the first five years of operation which
 includes \$5.0M for the
purchase of the required mining equipment. See Table~\ref{comparison}
 which compares
 single and three chamber construction costs.

\subsubsection{Continued Chamber Construction}

The next three chambers can be started, all at the same time, during
 the final year of 
work on the first three chambers.
With this excavation schedule, it is likely that installation of a
 Cherenkov detector
 in at least one chamber can begin four years after the start of the
 excavation process, 
with other installations following soon after, depending on the flow
 of funds and
 photomultiplier tubes (PMTs). The feasibility of doing much of the 
work to achieve 
completed physics modules is the result of the layout of the individual
 modules shown
 in  Fig. \ref{linout}, 
in which the spacing between adjacent modules,
450 ft., is sufficient to allow excavation of one chamber at the same time as
 work on
 liners or structures to support PMTs is carried out in other chambers.

In any event, we expect to have at least three chambers 
ready for experiment five years after the start of the program and 
perhaps  five,
modules ready for experiment at the end of seven years.

\section{Design and Cost of One Module of the Cherenkov Detector}
\label{onemodule}

 There exists considerable experience in designing 
and constructing the detection systems of large water Cherenkov
and other neutrino detectors: in 1978,
 at  BNL, a system with 13,000 proportional drift tubes (PDT) for tracking and 
4,000 
PMT for calorimetry \cite{734nim}; in 1983 for a long baseline 
neutrino experiment at BNL \cite{e776}; in 1986 for the
 Kamiokande-II detector in 
Japan, the electronics and DAQ for a system of 1,000 20 inch PMT \cite{kam2}; 
and in 1997 for 
the Sudbury Neutrino Observatory, a system of 9,500 PMT and  DAQ \cite{snonim}.
And of course, there is the accomplishment of Super-Kamiokande \cite{sknim}.

The Cherenkov detector to fill a chamber at Homestake is similar in its essentials to 
Super Kamiokande and SNO; it is 
composed of detecting
elements---photomultiplier tubes (PMTs); 
the PMT supporting mechanical 
structures;  associated electronic and DAQ
systems; the calibration systems; and the water purification
system. Here we outline the  technologies and show preliminary
cost estimates for each of these systems.
It is important to note that all of these technologies are well 
understood, there are no new or unusual developments required, and
the draft designs and costs are based upon experience with successful
prior  detectors.

\subsection{Photomultipliers}
The core of an imaging Cherenkov detector is the array of PMTs.
 The complexity and cost of the detector depend
largely on the total number of PMTs while the
the total photocathode coverage determines  
energy resolution as well as the energy threshold; the timing
performance and the density of the PMTs largely determines the reconstruction
accuracy. Therefore PMT selection is critical to the affordability
and to the sensitivity of the final detectors. Clearly much work needs to
be invested in the many details of actual PMT procurement, nevertheless it
is possible to make a reasonable estimate based upon experience and
recent contacts with possible manufacturers.

Because the cost per unit area of photocathode declines somewhat with
larger PMTs and the cost of many of the supporting systems like cables
and electronic channels declines directly with smaller numbers of 
PMTs, it is sensible to think of relatively large diameter PMTs.
The very largest devices, the Hamamatsu 50 cm tubes, have, however,
a significant mechanical disadvantage in terms of stored energy versus
envelope strength as demonstrated in the SuperKamiokande accident.
Therefore we have chosen to base our estimates on tubes of about
25 cm diameter. Hamamatsu, R7081, Electron Tubes, 9360KB, and Photonis, XP1807
photomultiplier tubes are
 in this class, all with specified performance superior to  
the tubes used in the previous generation of large water Cherenkov
detectors. For example, the XP1807 claims a quantum efficiency
of 24\% at 420nm and a Transit Time Spread of 2.4ns FWHM. Recent
measurements of prototype tubes at support these numbers.

The fine tuning  of the PMT coverage and the granularity, which determines 
the best size of the PMT, must await detailed Monte Carlo simulations. 
But based on simple estimates of light yield  and past experience we can make 
a good first design.   The energy threshold of the detector is determined 
by the total coverage. For 25\% coverage, a 5 MeV electron should give a
signal of $\sim 25$ photo-electrons.  This is sufficient for 
good detection efficiency if the radioactivity background 
can be controlled. 
The light yield  number will be lower for 
a 5 MeV gamma ray because of the energy lost in Compton scattering. 
The detection efficiency for $\sim 6$ MeV gamma is important for the 
proton decay mode $p \to k^+ \bar\nu$ tagged by a 6 MeV de-excitation gamma
of about 6 MeV.  Therefore the total photo cathode coverage of 
25\% or greater  is set by the 
need for a 5-6 MeV threshold. 
The granularity is set by the need for good pattern recognition to
reject background events with multiple particles and select good electron
neutrino interactions.   Work reported in \cite{chiaki} indicates that 
higher granularity will be important for this purpose. 
A 1-2 GeV muon will multiple scatter in water to  give an additional 1 to 
2 degree scatter to the Chrenkov photons. Therefore, the choice of 
PMT granularity to match this multiple scattering limit will give us 
the best possible performance for pattern recognition for 
this type of detector. This corresponds to spacing of about 1 meter for 
a fiducial radius of 25 m.

For this estimate we use a total photocathode coverage of 25\%
which implies  $\sim$50,000 PMTs per 100 kT module. The cost
used in the total PMT estimate is scaled from the 20 cm tubes used in SNO and
is  not based on manufacturer's quotations.  Nevertheless, all manufacturers 
have accepted, in principle, that the estimate of \$800 per PMT
is not unrealistic. To this \$800 number we add an additional 
10\% for 
the PMT testing costs plus
the cost of the base and the waterproofing that surrounds the base.

We make the further assumption that at the planned depth of the chamber it will
not be necessary to have a sophisticated anti-counter
 (see Figure \ref{fig1}) and  have allowed only for a few meters of water
at the top and bottom of the cylinder to be optically isolated
and equipped with a small number of anticounter PMTs looking
up or down. This is similar to the SNO approach where only 91 PMTs 
are used to observe the outer light water volume.

\subsection{Mechanical Structure}

The individual PMTs must be fixed in well known locations evenly 
distributed across the interior surface of the 50 m by 50 m cavity.
To do this, a mechanical structure must be designed and
installed.   There are a number of suggested
schemes for achieving this mounting, but not yet any engineered
plan. Therefore we  estimate this system cost on   general
grounds. The SNO mechanical mounting was for a spherical surface,
much more complex than the simple cylinder of this detector, and therefore
not a good cost model. The SK scheme is plausible for this case, but we
do not have access to their cost data. 
Including installation labor, a simple model gives about \$165 per
PMT total cost for the structure.

\subsection{Electronics and DAQ}

The basic SNO electronics system has performed nicely and that architecture
should work well for the detector proposed here. However, the custom integrated
circuits were manufactured in processes that are no longer available
and many of the commercial parts are obsolete or obsolescent so a
redesign in detail will be necessary. This will also present an
opportunity to remove some of the minor flaws of the SNO
implementation.

The SNO production and development costs are well documented and  
we have  scaled  for inflation to arrive at a per PMT
cost of about \$120 for front end electronics; trigger system; low
voltage power; and high voltage power. The development costs are taken
as the same number of people-years  required for the original SNO system
plus silicon chip and printed circuit prototyping costs.

The DAQ system involves largely development, but there is a small cost
for the computers needed for data recording; on line monitoring, for slow
controls (largely environmental controls); and miscellaneous networking
and GPS interfaces. These add a fixed cost per module of about \$75k.
%An additional central computing farm to support off-line analysis would
%probably be included in a full design report but is only mentioned
%here. 
The DAQ development costs are assumed similar to SNO's at about
five people-years.

\subsection{PMT Cables and Connectors}

The cable running from each PMT to the electronics, presumably located
around the periphery of the module above the water surface, is
 a significant cost item. For a 50 m diameter x 50 m high detector and equal
length cables to allow an isochronous trigger, each cable needs to
be at least 70 m long. High quality waterproof transmission line
similar to that used in SNO costs about \$0.50 per meter to which one
must add the assembly and testing costs and the connectors.

It is possible to imagine for the right circular cylinder geometry of
a  module that it would be possible to eliminate a {\em wet end}
connector and simply pot the cable and base as a unit. For conservatism,
and because there are unknown installation costs associated with pre-
attached cables, we have assumed a {\em wet end} connector with full
waterproofing for a total assembled cable cost of about \$77.

\subsection{Calibration Systems}

The absolute energy calibration in the GeV range 
is necessary for the program we have proposed here.
Super-Kamiokande has demonstrated energy scale calibration at 
$\sim 2.5\%$ \cite{sknim}. We would like to accomplish 
approximately a factor of 2 better than this.   
The energy calibration in the MeV range is also important for 
other physics.  
A number of levels of calibration are required of an imaging
water Cherenkov detector. The most basic level is  a series
of electronic calibrations of charge and time 
in well defined units, which  is included in
the electronics costs above. However, it is also necessary to have an
optical calibration system to define a common time reference and to
measure the optical properties of the detector. We choose a fast,
multi-wavelength laser source and a diffusing ball capable of being
moved
through a large fraction of the total detector volume. Finally, one
needs  handles on actual energy scales. The  peak at the end point of 
stopped muons is 53 MeV.
At least one other energy point is needed to give  the
slope of the energy calibration curve. It is 
 useful to consider the Super Kamiokande 
scheme of a small electron accelerator
to provide a  versatile calibration source---this is especially
attractive if it can be shared amongst all the modules.
 We include a laser system;
a source positioning apparatus; and a neutron generator and decay chamber
for one nuclear source as a development cost to arrive at a total
of about \$400k for calibration systems.

\subsection{Water Purification and Cooling}

The water system is modeled on the SNO light water
system which has produced water of exceptional cleanliness and
low radioactivity. For costing purposes we have simply scaled the
SNO costs by one half the volume ratio on the assumption that we would
require the same level of transparency as SNO - greater than 200 m in the
region between 350 and 450 nm (very near the Rayleigh limit). This
implies that we would want to have the same level of sophisticated
reverse osmosis and filtering equipment 
and  would want to maintain nearly the same rate of turnover.
%However, the full volume ratio is clearly an overestimate as fourteen
% three inch pipes cost much more than 
%one twelve inch pipe and fourteen small pumps are more expensive than
%one or two much larger pumps. 

In addition to the purification system, we also require the water to
be degassed to inhibit biological activity and cooled to further inhibit
biology and to keep the PMT noise rates low. The degasser is included in
the SNO costs as are the  heat exchangers. As the SNO
costs also include plumbing from a (closer) chiller station, we have
simply scaled those costs by our factor of seven (one half of the volume
ratio of 14) to get \$4.5M. A full design effort will 
probably reduce these estimates.

\subsection{Summary of Detector Module Costs}

\paragraph{Development Costs} for the major detector systems
are extrapolated largely from the SNO cost history with which
we are well acquainted. These costs, which will be shared among
three modules, are estimated at \$3.0M including inflation. Some
of the tasks require professional engineering while some (e.g. DAQ
systems) are largely physicist tasks.

\paragraph{Procurement Costs} for module hardware we estimate \$5M,
\$4.5M is for the water purification and distribution system,
and \$0.4M for calibration. 
%Note that the cost of the electron
%accelerator for calibration is shared amongst all three modules.

\paragraph{Production Costs} for the items directly in the photomultiplier
chain total approximately \$1,242 per PMT. This estimate includes assembly
and quality assurance labor as well as the final installation labor. For 25\%
PMT coverage of the total $11,000m^2$ area of a single module, the
number of PMT required is approximately 50,000 costing \$62.1M.

Thus for the three modules, we have a total detector 
cost, including all development, procurement and production costs, 
of about \$194.2M. Applying a 25\% contingency
to the non-excavation costs (a separate contingency was already applied to excavation) gives a total cost of about  \$242.7M.
 This cost is dominated by the per
PMT production costs. Clearly, it behooves us to do what can be done to reduce the cost per PMT. We have suggested to several PMT manufacturers that
potential sales of at least 150,000 PMT might pay for the construction
of a PMT manufacturing plant near the Homestake Mine. They responded that
this could result in substantial savings on component delivery costs,
testing and reworking of finished PMT, tax relief, etc. If funds for this
proposal were to become available, further discussion along this line
might proceed profitably.

\subsection{Final, Overall Cost of Three Detector Modules
 (including excavation
of underground chambers)}

From section ~\ref{designandcost}, the estimated cost of excavation
and completion of three underground chambers will be \$66M (including 30\%
cost contingency). Hence the total overall cost of the first three-tenths
of a megaton detector is \$308.85M, to be spent over five to six years in 
which the first 4 are for cavity construction.

\section{Broadband Neutrino Beam to DUSEL}

As mentioned above, the accelerator neutrino beam will be produced by
one of the hadron accelerating laboratories, BNL and Fermilab, 
and consequently will be their responsibility. Nevertheless, a few comments on the subject 
here may be useful.

\begin{figure}[htbp]
  \begin{center}
 \includegraphics*[width=\textwidth]{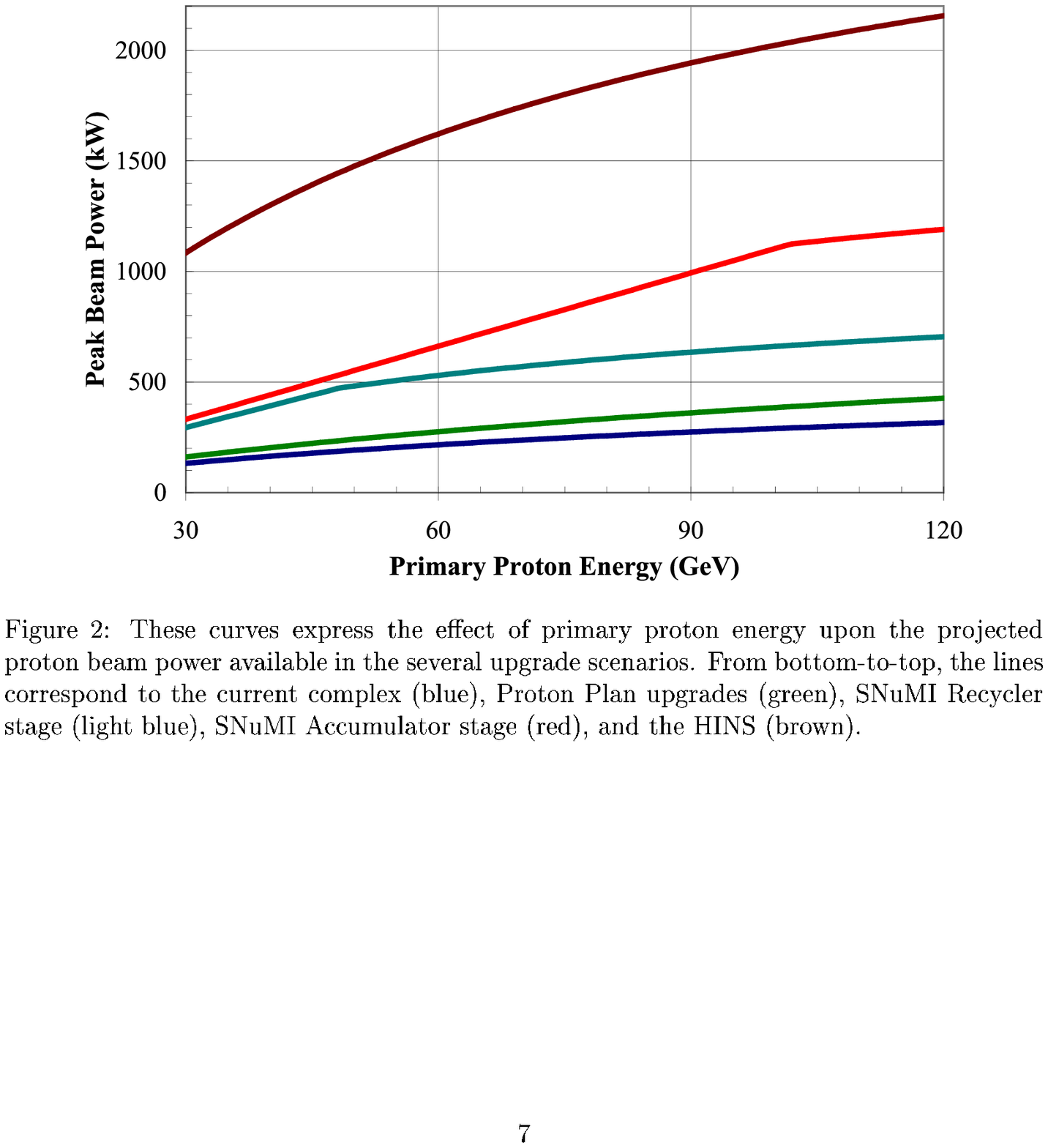}
  \caption[Preliminary proton power projections at FNAL]{
Preliminary estimates 
 of increasing proton beam power in kWatt as a function of 
proton energy at Fermilab. 

}
   \label{ppower} 
  \end{center}
\end{figure}

Our initial plan to study neutrino oscillations and search for CP violation
is based on a wide band neutrino beam generated by 30 to 60 GeV protons
incident on a target to produce pions and kaons focused by a magnetic horn;
such neutrino beams with a power between 0.3--0.7 Megawatt could be 
available at both laboratories. A plan to 
re-steer the proton beam, provide a new horn and target system, 
and a new meson decay path has been developed
at BNL (for the cost and schedule see \cite{agsup}). 
A   plan to create a beam at FNAL is being devised at part of the joint 
FNAL/BNL effort. Some initial results of these studies are in 
figure  \ref{ppower}   \cite{zwaska, bock}.
The beam could be built by reusing the existing extraction for NUMI and directing the proton beam towards the west instead of the north. 
It has been determined that there is sufficient space on site at FNAL to 
accomodate a 380 meter long decay tunnel and a near detector.  
The beam from either laboratory will yield
 data on neutrino oscillations and thus information on the neutrino
mass hierarchy, on the mass effect and on $\delta_{CP}$ will be obtained,
as outlined in Figs. \ref{fig2} to \ref{antiapp}. As always, atmospheric neutrino data
and proton decay data will also go forward.

Clearly, increased neutrino beam power will be an advantage in obtaining
measurements of neutrino oscillations and of $\delta_{CP}$.
 This can be accomplished with high proton intensities on the target,
but a target able
to sustain higher proton intensity must  be developed. Targets of 2 
megawatts are curently contemplated and there is active joint FNAL/BNL effort
to understand the properties of possible new target materials and geometries.

\begin{figure}[htbp]
  \begin{center}
 \includegraphics*[width=\textwidth]{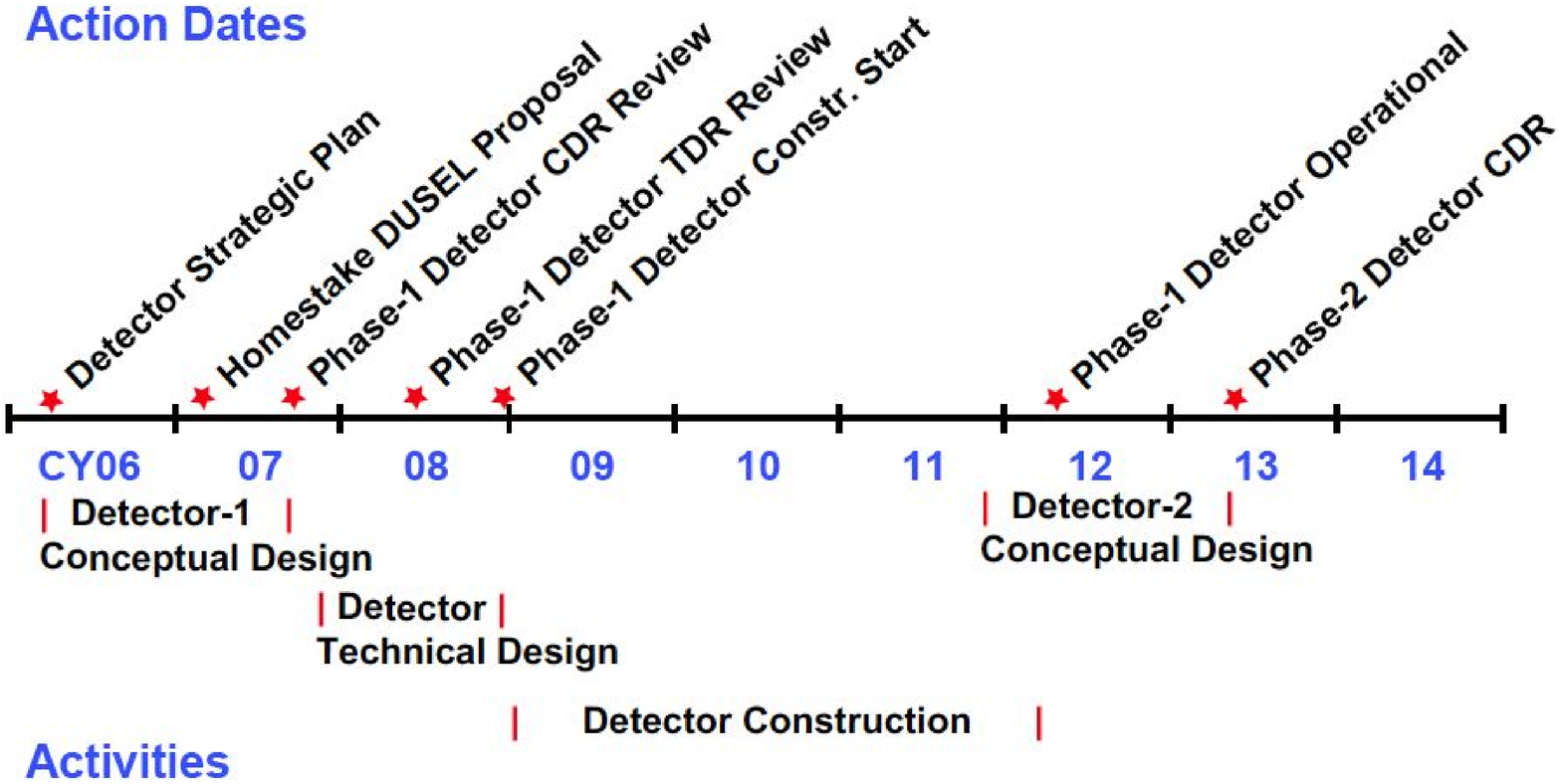}
  \caption[Possible timeline for the first phase of 
construction]{
Possible timeline of reviews and funding decision for construction 
of the first 3 modules of the detector array.  
}
   \label{tomtime}
  \end{center}
\end{figure}

It is not necessary or desirable, however, to delay the start of construction
of the 300 kiloton neutrino detector in DUSEL until that time because
construction of both the detector and the more intense neutrino beam
can proceed simultaneously and independently during the next few years. 
In Fig. \ref{tomtime} we have produced a possible aggressive time line for 
proceeding with the construction of the first 3 modules of the detector 
array presented here. 
A proposal for a joint
 effort by BNL and Fermilab to accomplish this task is being
studied at present and should be available before the end of 2006

%\begin{figure}[htbp]
%  \begin{center}
% \includegraphics*[width=\textwidth]{fnal-beam.eps}
%  \caption[DUSEL neutrino beam from FNAL]
%{
%Preliminary drawings of locating a 400 km 
%long decay tunnel for a new neutrino beam to DUSEL 
%at either Homestake or Henderson facilities.    
%
%}
%   \label{fnaldusel}
%  \end{center}
%\end{figure}

\section{Summary}

As mentioned at the beginning, this document is the first part---the
 conceptual design---of a formal proposal to construct a 
key 
shared national physics
 research facility
in the Homestake Laboratory.
 It will carry out a program of important
experiments in neutrino physics and other elementary particle physics.
 The proposal described here is primarily for a program of measurements 
 to be done 
in the facility utilizing neutrino beams far from their accelerator origin
to study fundamental features of the lepton sector of elementary particles.
The laboratory and the experimental apparatus constitute the basis of a plan
of considerable ambition and cost, but all aspects of the plan---including
that for the accelerator-generated neutrino beams---have survived severe
scrutiny concerning scientific potential, feasibility, and cost.

We have attempted  to indicate an approximate 
schedule of events in the progression from the Homestake Mine to 
the Homestake Laboratory to the construction of a massive neutrino 
detector in the Laboratory. Our vision of that schedule is summarized in 
Fig. \ref{tomtime}.

The other measurements and searches of importance---including the search for
proton decay---will be done independently and at the same time that the
neutrino oscillation and CP-invariance measurements are in progress. The former
do not require an accelerator-generated neutrino beam. They are discussed
here briefly for completeness, but will be covered in more detail in
the second part---the technical design---of the proposal. The present part
concentrated on the overview of the program, and 
 the array of imaging water Cherenkov counters; those
counters and their underground accommodations, which are described at length
here, are at the heart of the proposal.

The difficulty of the proposed experiments is surpassed only by the importance
of the questions they seek to answer. These demand high statistical precision
and high signal to background discrimination to guarantee clarity
and unambiguity in the interpretation of the results.

The quality of the facility at the Homestake Laboratory and the opportunity
to use it to pursue fundamental physics for many years will make
Homestake one of the dominant centers of experimental particle physics
in the world.

\newpage 

\appendix{\bf Appendix 1.  Further comments on 
preservation of fiducial volume and 
 optimizing the size of the cavern at 4850 ft} 

Added on June 30, 2006 after comments from the NUSAG committee.

The fiducial volume for the detector will clearly depend on 
the eventual optimization of both the photomultiplier 
 hardware structure as well 
as the software cuts for reconstructing the events of interest.  
It also depends on the scientific focus of the detector. 
Before expanding on the ideas we will first examine what was done for 
 Super Kamiokande \cite{sknim}. 
Super Kamiokande cavity has a stainless steel tank that is self-supporting 
with concrete backfilled against the rough-hewn stone walls. 
The dimensions of the tank are 39 m in diameter and 42 m high, corresponding to 
$>$50.0 kTon mass.   Inside this tank is a cylindrical PMT support structure 
with inner dimensions of 33.8 m diameter and 36.2 m high.   This structure 
contains $\sim$32 kTon of water and it divides the volume into an inner 
detector and an outer detector, which is used both as a veto for incoming particles
and a passive shield against neutrons and gammas from the rock. 
Such a passive veto volume is necessary for suppressing background to 
 very low energy events  from solar neutrinos. 
For the reconstruction of most physics events the Super 
Kamiokande collaboration 
makes the requirement that the event vertex be $>$2 m away 
from the wall of the 
inner detector. This final requirement brings  the fiducial volume to 
$\sim$22 kTon.  

Although, in much of the proposal above we have made detailed 
estimates for a 50 m diameter and 50 m high volume of water and the 
cavity that holds it, we are cognizant of the fact that the 
exact dimensions of 
the cavity will depend on a detailed  optimization concerning 
the thickness of the veto 
volume, and the dimensions of the phototube mounting apparatus. 
Our intention is to preserve the 100 kT fiducial as well as we can without
creating engineering requirements that are more difficult than the ordinary.

We were surprised in the interest in the fiducial volume of our
detector modules and the exact mounting location of our
photomultiplier tubes.  
The plan for the excavation presented above will create stable 
water-tight cavities, 50 m in diameter and 50 m high
 that have a good inner surface lined by concrete. 
We are focused on 11 inch, 28 cm, diameter
tubes such as Electron Tube 9360KB or Photonics XP1807.  Both of these
tubes are approximately 30 cm long, from the peak of the photocathode
window to the base.  We are considering supporting the tubes at their
widest point about 12 cm from the peak of the photocathode window or
18 20 cm from the base.  Connecting the mount at this position rather
than at the base eliminates the torque at the tube neck associated
with the buoyant force on the spherical region of the tube and the
lever arm to the tube base.  That is particularly significant for
tubes near the bottom of the detector, 50 meters below the water
surface or at a pressure of 5 atmospheres.

Our preliminary plan is to have the peak of the photocathode window
about 50 cm from the chamber wall.  That puts the PMT support point
about 38 cm from the chamber wall.  We could easily attach a TYVEK
partition to the support structure and thus create a 38 cm wide,
optically isolated veto region as suggested by Peter Meyers, without having
any effect on the fiducial volume.  A similar optically isolated
region at the top and bottom are easily achieved without any effect on
fiducial volume.

For long baseline physics we are interested in relatively high energy events
($>$ 500 MeV). For such events it might be sufficient to have 
a relatively thin veto region and 
an aggressive fiducial cut  with a  boundary  1 meter inward from the
photocathode window  or 1.5 meters from the cylinder wall.  
Clearly this must be confirmed by detailed simulations that will be performed 
as we design this detector. Nevertheless, 
such a cut leaves a 47 meter diameter fiducial volume.  In order to maintain a
100 kiloton fiducial volume, that requires a 57 meter fiducial region
height instead of the stated 50 meter height. 
In the cavity design there is much lattitude in  the vertical direction. 
 For the bottom we have about 15 meters of water beneath the lower PMT
 layer that we can use as a veto region.  On the top, we merely have to
 lower the PMT layer by 30 - 50 cm to make the veto region.  
 All this requires is
that the bottom layer of PMT be a bit deeper.  The issue here is that
the PMT manufacturers may be reluctant to have their PMT under more that 5
atmospheres pressure, especially after the Super Kamiokande accident.
We hope to explore this matter further as soon as we have some funds
with which to carry out PMT envelope pressure tests.

A second option that maintains the 100 kiloton fiducial volume is to
expand the excavated cylinder by 1.5 meters in radius and keep the
bottom PMT at a 50 m depth. This is probably a safe procedure
especially since the initial stress evaluations were carried out for
detectors at 6950 ft depth rather than at 4850 ft depth.  This matter
should be reconsidered after we have coring data and rock
characteristics in the excavation region.  As  mentioned above,
our goal is to obtain these cores in early 2007.

If we keep the chamber dimensions exactly as stated, 50 meter diameter
and 50 meter diameter, then the reduction in fiducial diameter to 47
meters will result in a fiducial mass of 86.7 kilotons per module or
260 kilotons for the three module complex.  That means a data rate
that is reduced by 13%.
If we are forced to make a 2 meter fiducial cut (such as Super Kamiokande)
  from the PMT window 
then the reduction will be to 45 m in diameter leading to fiducial mass of 
71.5 kilotons per module or 214 kilotons for the three modules. 
This is still a less than 30\% reduction in data rate which will not have 
a major impact on the scientific case for the program.  

We anticipate that we will find a suitable combination of chamber
radius and bottom PMT location that will achieve the  100
kiloton fiducial volume per module.  As stressed before,  we
are very concerned that there be large safety factors in the
excavation structure and the PMT installation.  Our goal is to
construct detectors that will have close to a 100\% on time efficiency
and so make maximal use of the long range neutrino beam as well as
maintain a constant search for nucleon decay, supernova neutrino
bursts and measurement of atmospheric neutrino behavior.

The 100\% on time is further aided by having a multi-module detector
array.  For example, we can stagger the module calibrations in such a
way that two modules are operating while the third is being
calibrated.

\end{document}